\documentclass{aa}

\usepackage{color}

\usepackage{graphics}
\usepackage{amssymb,amsmath}
\usepackage{amsmath}
\usepackage{psfrag}
\usepackage{graphicx}

\usepackage{float}
\usepackage[caption = false]{subfig}

\usepackage{graphics}
\usepackage{amssymb,amsmath}
\usepackage{amsmath}
\usepackage{psfrag}
\usepackage{graphicx}
\usepackage{natbib}

\newcommand{\ve}[1]{\mathbf{q1}}

\newcommand{\be}{\begin{equation}}      
\newcommand{\ee}{\end{equation}}      
      
\newcommand{\bef}{\begin{figure}}      
\newcommand{\eef}{\end{figure}}      
\newcommand{\bea}{\begin{eqnarray}}    
\newcommand{\eea}{\end{eqnarray}}

\begin{document}

\title{Nonaxisymmetric models of galaxy velocity maps} 
  
\titlerunning{Nonaxisymmetric models of galaxy velocity maps}
  
\authorrunning{Sylos Labini et al. }  
  
  \author  {Francesco   Sylos  Labini  \inst{1,2,3},   David  Benhaiem
    \inst{2},   S\'ebastien   Comer\'on   \inst{4}  and   Mart\'\i   n
    L\'opez-Corredoira\inst{5,6} }

        \institute{Museo  Storico  della  Fisica   e  Centro  Studi  e
          Ricerche Enrico  Fermi, I-00184 Rome, Italy  \and Istituto dei
          Sistemi Complessi, Consiglio Nazionale delle Ricerche, I-00185
          Roma,  Italy  \and   Istituto  Nazionale  Fisica  Nucleare,
          Dipartimento  di  Fisica, Universit\`a  ``Sapienza'',  I-00185
          Roma,  Italy \and  University of  Oulu, Astronomy  Research
          Unit,  P.O. Box  3000, FI-90014,  Finland \and  Instituto de
          Astrof\'\i sica  de Canarias,  E-38205 La  Laguna, Tenerife,
          Spain \and  Departamento de Astrof\'\i sica,  Universidad de
          La Laguna, E-38206 La Laguna, Tenerife, Spain\\}

\date{Received / Accepted}

\abstract{Galaxy velocity maps{  often show} the typical
  pattern of a rotating disk,  consistent with the dynamical model where
  emitters rotate in circular orbits around the galactic center.
  The simplest template used to fit these maps  consists in  {  the rotating
   disk} model (RDM) where the amplitude of circular
  velocities is fixed by the observed velocity profile along the
  kinematic axis.  A more sophisticated template is the rotating
  tilted-ring model (RTRM) that {  takes into account the presence of warps and} 
  allows a radius-dependent orientation
  of the kinematic axis.  In both
  cases, {   axisymmetry is assumed and} residuals between the observed and the model velocity fields
  are interpreted as noncircular{   motions.}
  We show that if{  a}  galaxy is not axisymmetric, there is an
  intrinsic degeneracy between a rotational and a radial velocity
  field.
  We then introduce a new galaxy template, the radial ellipse model
  (REM), that is not axisymmetric and has a 
    purely radial velocity field 
     {  with an amplitude}  that is correlated with the major axis of the ellipse.
  We show that best fits to the observed two-dimensional velocity
  fields of 28 galaxies extracted from the THINGS sample with both the
  REM and the RDM give residuals with similar amplitudes, where the REM
  residuals trace {  nonradial motions}.
  Best fits obtained with the RTRM, because of its larger number of
  free parameters, give the smallest residuals: however, we argue that
  this does not necessarily imply that the RTRM gives the most
  accurate representation of a galaxy velocity field. Instead, we show
  that this method is not able to disentangle between circular and
  radial motions for the case of nonaxisymmetric systems. 
  {    We then 
  discuss a refinement of the  REM, able to describe the properties  of 
  a more heterogeneous velocity field where circular and radial motions 
  are respectively predominant at small and large distances from the galaxy center. }
  Finally, we{  consider} the physical motivation of the REM, 
    and discuss how the interpretation of galactic dynamics
  changes if one assumes that the main component of a galaxy
  velocity field is modeled as a RDM/RTRM or as a REM. }
  
  \maketitle


\section{Introduction}

Two-dimensional velocity maps of {   many} galaxies show the
distinctive pattern of a rotating axisymmetric disk, that is, the typical
velocity gradient where on one side of the nucleus spectral lines of
stars (or other emitters) are shifted toward the blue region of the
spectrum with respect to the systemic velocity and on the opposite side
lines are shifted toward the red spectral region \citep{Rubin_1983}.{  These observations are}  usually interpreted as originating from the Doppler
shift caused by the circular motion of the various emitters around
the center of the galaxy.  The dynamical model that is derived from
these data postulates that {  a }  galaxy is close to a steady rotating
 axisymmetric disk configuration in which centripetal and centrifugal
forces compensate each other at all radii.  By comparing the
line-of-sight velocity profile with the amount of luminous matter it
is concluded that, in order to maintain such a steady state, a large
amount of dark matter is then needed
\citep{van_der_Kruit+Bosma_1978,Thonnard_etal_1981, Bosma_1981}.  In
particular, evidence for the existence of dark matter halos around
spiral galaxies comes mainly from the flatness of the rotation curves
outside the visible region of galaxies with the extended HI emissions
\citep[see][for a review]{Sofue_Rubin_2001}.

Coherently with this model, observed galaxy velocity maps are usually
fitted with a template consisting of a{   rotating  disk}
model (RDM): this assumes that a disk {  (axisymmetric)} 
galaxy is in circular rotation
in a plane about a central axis. The amplitude of the circular
velocities as a function of the distance from the center, that is, the
rotation curve, is obtained from the observed one-dimensional (1D)
line-of-sight (LOS) velocity profile measured along the galaxy
kinematic axis\footnote{The kinematic axis is the axis passing through
  the center of mass of the distribution and along which the
  difference of the observed velocities at the two extreme points is maximal.}
\citep{Begeman_1989,Schoenmakers_etal_1997,Beckman_etal_2004,Trachternach_etal_2008,Erroz-Ferrer_etal_2012,Erroz-Ferrer_etal_2015}.
The best-fitting RDM is the one that minimizes the residuals between
the rotational model velocities, computed for a specific value of the
inclination angle\footnote{The inclination angle is the angle between
  the LOS of the  observer and a vector orthogonal to the plane of the
  disk.} of the disk $i$ and the actual data. Significant residuals
are typically measured in such fitting procedures -- of the order of
20-30\% of the maximum circular velocity or even larger -- and
these are attributed to noncircular (e.g., radial, random, etc.)
motions \citep[see,
  e.g.,][]{Jorsater+vanMoorsel_1995,Zurita_etal_2004,Trachternach_etal_2008,Sellwood_etal_2010,Erroz-Ferrer_etal_2015}.

The RDM is only compatible with observations at first order: for
instance in the case of an ideal rotating disk, by construction, the
kinematic axis must be aligned with the projected semimajor axis,
whereas it is frequently observed that galaxies show a significant
angular offset between these two axes \citep[see,
  e.g.,][]{Erroz-Ferrer_etal_2015}.  In addition, it is known that
many galaxies exhibit bars and/or warps that can locally distort the
velocity field and that cannot be described by{  the} simple axisymmetric
disk model. Indeed, several observations have shown that most disks
exhibit a wealth of nonaxisymmetric structures
\citep{Rix+Zaritsky_1995,Kornreich_etal_2000,Laine_etal_2014} and that
the stellar disk in a typical spiral galaxy is significantly lopsided,
indicating asymmetry in the disk mass distribution. 
Lopsidedness is quite typical in disk galaxies \citep{Jog+Combes_2009}
and {  it} may be interpreted as a pattern of elliptical orbits
\citep{Baldwin_etal_1980, Song_1983}.

{  A simple disk is clearly not a realistic representation of a galaxy, but 
introducing a more complex shape is very difficult and requires 
modeling systems of increasing complexity. In this respect a 
relatively simple way to} 
take into account the fact that a galaxy disk 
{   may exhibit warps was to introduce} 
 the rotating tilted-ring model (RTRM).  
{  In particular,  the physical motivation to 
hypothesize this template was to 
accommodate warps in HI disks originally detected for the case of M83 
\citep{Rogstad_etal_1974}. 
In that case it  
was indeed observed that the velocity field 
was  incompatible with a simple RDM and it was 
thus proposed to 
interpret the observations in terms of a “warped
disk” where all the mass moves in circular rotation
around the galaxy center, but where the material that
lies beyond the optical image moves in orbits inclined with respect 
to the central plane. More specifically, the inclination of the orbits 
was thought to depend on the distance from the galaxy center. In this way the RTRM,}
similarly to the RDM, assumes axisymmetry but
it also postulates that a galaxy can be described as a set of
concentric rings where each ring is characterized by a circular
velocity and an orientation (see below).  Changing the orientation
angles as a function of the distance from the galaxy center
{  in a continuous way} makes it
possible to obtain a better fit than with the RDM.  As for the case of
the RDM, the RTRM residuals are interpreted as the signature of
noncircular motions.  A great effort is then devoted to characterizing
residuals (i.e., the difference between the actual galaxy velocity
field and that of the best-fit RTRM) that are interpreted to trace
motions{   deviating}  from {  purely} rotational ones
\citep{Trachternach_etal_2008,Oh_etal_2008,Erroz-Ferrer_etal_2015}.

{   The existence of warps 
has been proven independently from kinematic
  studies, namely by observing both edge-on galaxies
  \citep{sancisi_1976,Reshetnikov+Combes_1998,Schwarzkopf+Dettmar_2001,Garcia_Ruiz_etal_2002,Sanchez-Saavedra_etal_2003}
  and the Milky Way 
  \citep{Levine_etal_2006,Kalberla_etal_2007,Reyle_etal_2009}. 
  These observations 
  give  a  straightforward physical explanation for
  a twisted position angle that justifies,
  from the physical point of view, the use of the RTRM. Of course 
  the presence of a warp is compatible with the RTRM but it does not
  prove either that a galaxy is axisymmetric or that emitters 
  move on stable circular orbits: these are however the two
  assumptions
  that are at the basis of the RTRM.}

In this paper we study the determination of a galaxy velocity field,
and in particular the problem of disentangling radial from circular
motions, in the case where the assumption of axisymmetry is not valid
to describe the shape of a galaxy.  We show that it is possible to { 
  build} a simple template that is very different from a RDM but that
fits the observed galaxy two-dimensional (2D) maps equally well (but both
models give worse fits than the RTRM). This template, referred to hear as
the radial ellipse model (REM), (i) breaks axisymmetry, i.e., it is an
(infinitely thin) ellipse, and (ii) has a purely radial velocity
field directed outwards that (iii) has a strong correlation with the
direction of the major axis of the  system.  We use the three models (i.e.,{  the} RDM,{  the} RTRM, and{  the} REM) to fit the
2D velocity maps of the galaxy measured by The HI Nearby Galaxy Survey
(THINGS) \citep{Walter_etal_2008}; we then compare the results between
the fits and discuss the different interpretation of the galaxy
velocity fields and dynamical models in the different cases.  In
particular, by considering the properties of some toy models with
physically motivated and complex velocity fields, we show that the
better fit typically provided by the RTRM{  does} not necessarily
correspond to the best representation of a given velocity field. Most
notably we show that this method, as the RDM, may confuse rotational
and radial motions if the system is not axisymmetric.

The paper is organized as follows: in Sect.\ref{sec:fits_sims} the
properties of some simple toy models allow us to illustrate the
problems encountered in disentangling the different motions (i.e., radial and
circular) in{  an ideal} galaxy velocity map if the assumption of
axisymmetry is not valid.  In Sect.\ref{sec:template} we
introduce the REM, discussing its properties and the various
parameters used in the fitting procedure.  We also detail the fitting
procedures of the three different models.  We then present in
Sect.\ref{things} the results of the fits with the RDM, with the REM
and with the RTRM of a sub-sample of 28 galaxies extracted from the THINGS
sample; {  we  also consider a template that consists in 
a combination of the RDM and the REM. 
We then illustrate the physical motivation of the REM and, finally,} 
in Sect.\ref{sec:dicus_concl}, we draw our main
conclusions, discussing the consequences of the breaking of axisymmetry on the
interpretation of galaxy dynamics and the estimation of the mass of a galaxy.


\section{Circular and radial motions in nonaxisymmetric objects}
\label{sec:fits_sims}

The observed 2D velocity field of a galaxy corresponds to
the projection onto the sky of a 3D one, where
measurements always give only the radial component of {  the } velocity 
{  of an emitter} in the
direction of the observer, that is, the LOS velocity.  By modeling a
galaxy as a disk {  (that is obviously axisymmetric)}, the projected LOS
velocities\footnote{with respect to the systemic velocity of the
  galaxy} can be written as (see, e.g., \cite{Begeman_1987,Beckman_etal_2004})
\be
\label{eq:beckman} 
v_{los} (r, \eta) = v_\theta \sin(i) \cos (\theta) + v_R \sin (i) \sin
(\theta) \;, 
\ee 
where, following standard conventions, $i$ is the inclination angle of
the observer, i.e.,  the angle between their LOS and a vector orthogonal
to the plane of the galaxy, $r$ and $\eta$ are polar coordinates (with
the angle $\eta$ defined relative to the axis orthogonal to the
observer LOS) in the plane of the sky of a point with coordinates $R$
and $\theta$ in the plane of the galaxy, and $v_\theta$ and $v_R$ are the
components of the velocity field, tangential and radial, respectively,
given in polar coordinates $(R, \theta)$ in the plane of the
galaxy. The polar coordinates in the two frames of references are
related by the transformation
\bea
\label{eq:beckman-tran} 
&&
\tan (\theta) = \tan (\eta) / \cos(i)
\\ \nonumber &&
R=r \cos (\eta) / \cos (\theta) \;. 
\eea
If the system has purely circular velocities and is axisymmetric
then $v_R=0$ and $v_\theta= v_\theta (R)$.  {  On}  the other hand, if the
system has purely radial velocities and is axisymmetric, then
$v_R=v_R(R)$ and $v_\theta= 0$.  In the first case the kinematic axis,
that is,  the axis passing through the center of mass of the distribution
and along which the difference of the velocities at the two extreme
points is maximal, must strictly be the major axis of the projection
for the case of a disk, while this is generally not the case for{  systems}  that are not axisymmetric (as we illustrate below).
Analogously, if there are only radial velocities, the kinematic axis
is orthogonal to the major axis of the projected image only for the
case of a disk.
 
In order to show the problems encountered in determining the respective contribution
of radial and circular motions in the case of a nonaxisymmetric{  system} that has a complex velocity field, let us consider a few
simple toy models.  We generate a toy model in three dimensions,
fixing its shape, that is, choosing whether it is a disk or an
ellipse\footnote{In reality it is a 3D object with
  thickness much smaller than its main linear dimensions.}. We then
determine its projection onto the sky of a random observer that is
identified by the inclination angle $i$ and by the azimuthal angle
$j$, that is, the angle between the projection onto the toy-galaxy plane
of the LOS and its 3D major axis (see
\cite{Benhaiem+Joyce+SylosLabini_2017} for details).

\begin{figure}
\includegraphics[width = 3.2in]{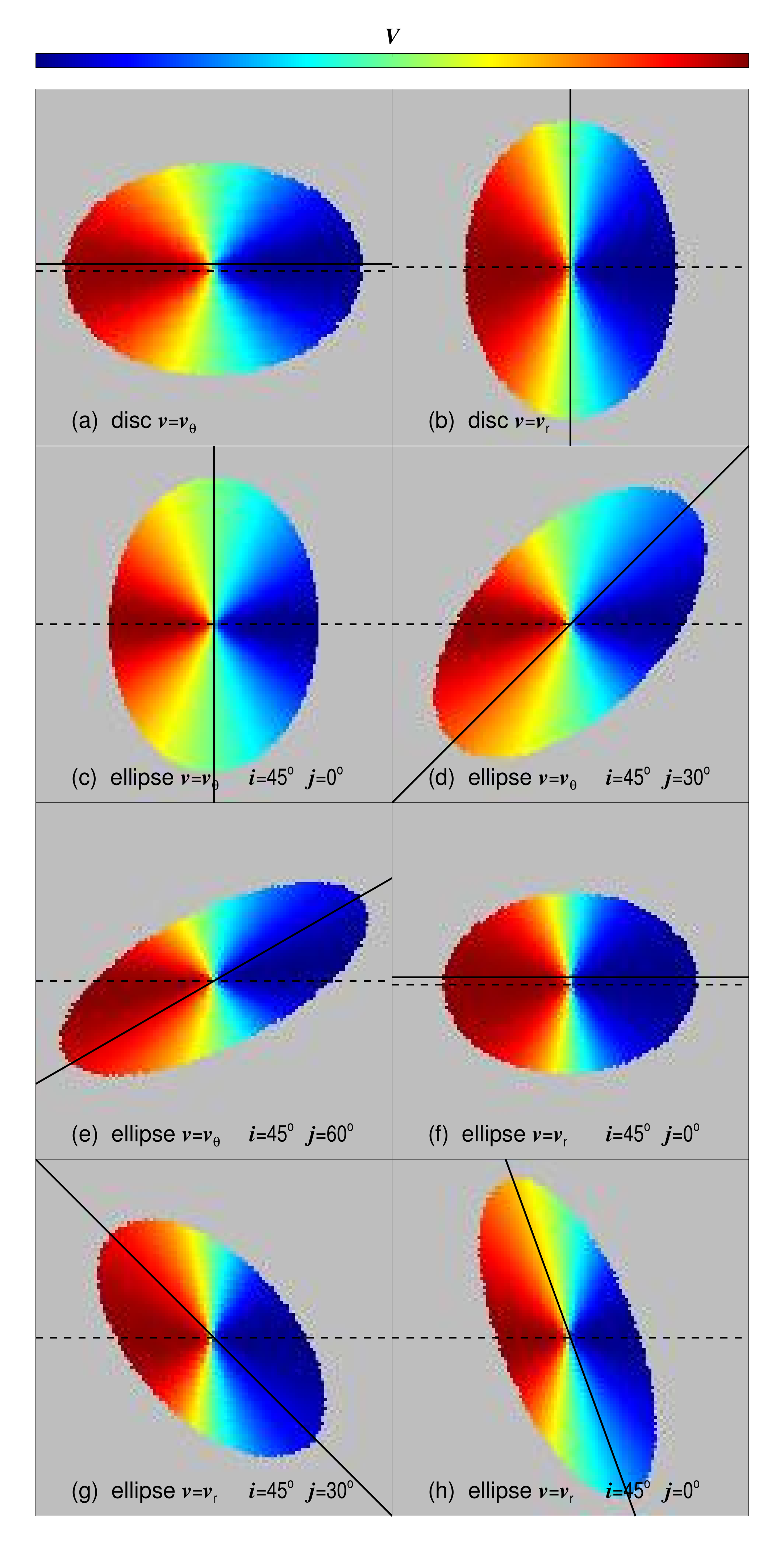}
\caption{  Projected velocity field for some toy models with angle $i,j$
  (the kinematic axis is shown as a dashed line, the projected major
  axis is a solid line).
  (a) Projection onto the plane of the sky of an
  observer with ($i=45^\circ, j=0^\circ$) of disk with purely circular
  motion.
  (b) As in (a) but for a  disk with purely radial velocities. 
   (c) As in (a) but for an ellipse with purely circular motions.
   (d) As in (c) but for $i=45^\circ, j=30^\circ$.
   (e)  As in (c) but for $i=45^\circ, j=60^\circ$.
   (f) As in (c) but for an ellipse with purely radial motions.
   (g) As in (f) but for $i=45^\circ, j=30^\circ$.
   (h) As in (f) but for $i=45^\circ, j=60^\circ$.
  }
\label{toymodels} 
\end{figure}

Let us start from the simplest case of a disk with purely solid-body
circular velocities: the projection, for $i=45^\circ$ and $j=0^\circ$,
is shown in Fig.\ref{toymodels}(a) where one may notice that the
kinematic axis and the projected major axis coincide one with the other
as predicted by Eq. \ref{eq:beckman}.  Figure \ref{toymodels}(b) shows the
projection of a disk with purely radial velocities directed
outwards\footnote{Of course the observed velocity map is symmetrical
  with respect to a change of sign in the velocity field: i.e., if we
  take radial velocities directed inwards rather than outwards
  observationally the system is the same modulo a rotation of
  180$^\circ$. The same occurs for all toy models discussed below and
  also for the REM. However, from a physical point of view a radial
  velocity field directed outwards is expected in some models of
  galaxy formation as we discuss in what follows.}: as expected,
according to Eq.\ref{eq:beckman}, the kinematic axis is orthogonal to
the projected major axis.  Because of the symmetry of the  disk, in these
two cases, the angle between the kinematic axis and the projected
major axis does not depend on the angles $i,j$.

Let us now consider the case of an ellipse with purely solid-body
circular motions:  Figure \ref{toymodels}(c)-(d)-(e)  shows 
three examples with parameters  $i=45^\circ$ and $j=0^\circ,
30^\circ, 60^\circ$.  The ellipse has a flatness parameter
\be
\label{iota} 
\iota=\frac{a_{max}}{a_{min}}-1 \;,
\ee
where $a_{max}, a_{min}$ are the intrinsic major and minor axes: as an
illustrative case we take a{  relatively} large value of the
flatness parameter, that is,  $\iota=1$.  One may note that in this case
the projected major axis forms angle $\psi$ with the kinematic axis,
where $\psi=90^\circ$ for $j=0^\circ$; then it linearly decreases,
becoming $\psi=0^\circ$ for $j=90^\circ$.  In
Fig.\ref{toymodels}(f)-(g)-(h) the case of an ellipse with
purely radial motions is shown, for the same values of $\iota$ and $i,j$ as
before. In this case we find $\psi=0^\circ$ for $j=0^\circ$; then
$\psi$ linearly increases with $j$, up to $\psi=90^\circ$ for
$j=90^\circ$.
These simple exercises show that the relation between the kinematic
and projected major axis, which occurs for a rotating disk, changes
 when the shape of the object is an ellipse; in particular,
for an ellipse, $\psi$ depends on the value of the angle $j$ 
{  and the functional behavior of $\psi(j)$ 
is relatively simple.} 

We can now introduce an additional, and crucial, feature of the
velocity field. Indeed, in a physically motivated model of galaxy
formation (see discussion below) it is quite natural that radial
velocities are oriented outwards and are correlated with the
major axis of the system.  For this reason the kinematic axis is aligned with the
projection of the major axis, which typically forms a small angle
(i.e., $\psi \ll 90^\circ$) with the projected major axis.  Thus,
contrary to an ellipse with purely radial velocities, in this
situation we expect that the kinematic axis and the major axis of the
projected distribution forms, even for large values of $j$, a
relatively small angle $\psi$, i.e., up to some tens of degrees.

\begin{figure}
\includegraphics[width = 3.5in]{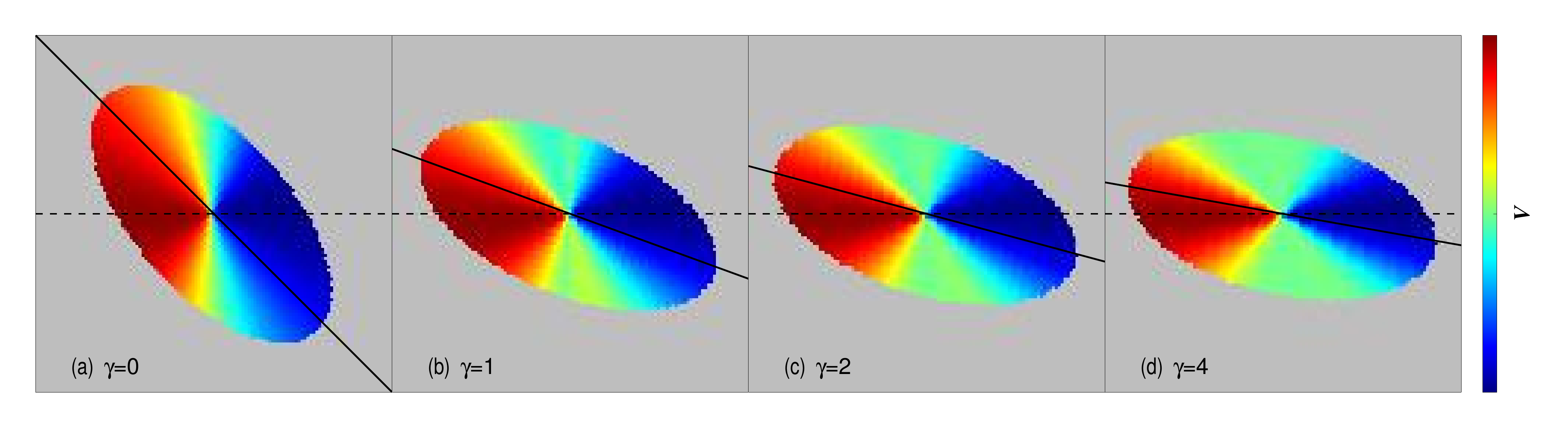}
 \caption{  Example of the angle between the projected major axis
 and the kinematic axis for an ellipse with $\iota=1$,   $A(r)=1$  
 for different values of the correlation exponent $\gamma=0,1,2,4$ 
 (see Eq.\ref{corr_vel})
  for the case $i=45^\circ$ and $j=45^\circ$ (see text for details).}
\label{toymodels2} 
\end{figure}

Let us therefore consider a simple toy model that presents such a
correlation.  For instance, one possible way to assign this kind of
nontrivial radial velocity is by fixing{  the 3D
  velocity field as}
\be
\label{corr_vel}
\vec{v(\vec{r})}= A(r)  \cdot \left| \cos(\omega ) \right|^\gamma \cdot \frac{\vec{r}}{|\vec{r}|}
,\ee
where $\gamma$ is an exponent that describes the strength of the
correlation between radial velocities and the major axis of the ellipse, $\omega$ is
the angle between $\vec{r}$ and the major axis, and $A(r)$ is a
function that describes the behavior of radial velocities as a
function of the distance from the center.  Figure \ref{toymodels2} (upper left panel) 
shows
the projection of{  such} a toy model with $\iota=1$, $\gamma=2,$ and $A(r)=1$
for certain values of the angles $(i,j)$: one may note that, as
expected, the angle $\psi$ between the kinematic and the projected
major axis is  small and the same occurs for other values of $(i,j)$.
{  In order to investigate the behavior of the angle $\psi$
as a function of the various parameters of this
  simple toy model, we have done several tests considering different values of 
  $\iota\;, A(r)$ and $\gamma$ and by considering several projection angles $i,j$. 
 An example for $\iota=1$, $A(r)=1$, $\gamma=0,1,2,4$, 
 $i=45^\circ$ and $j=45^\circ$ is shown in Fig.\ref{toymodels2}:
 we find $\psi \approx 45^\circ$ for $\gamma=0$ (i.e., no correlation)
and then it decreases when $\gamma$ grows up to 
 $\psi \approx 10^\circ$ for $\gamma=4$. Indeed, in the former case
 the alignment between the projected major axis and the kinematic axis 
 is a consequence only of the ellipsoidal shape of the system.
 Instead, when $\gamma$ grows the correlation between the direction
 of the radial velocity and the direction of the major axis 
 gets stronger and therefore $\psi$ decreases. Different values of the flatness parameter 
 and of the projection angles change the value of $\psi$ but not the trend with 
 the correlation exponent $\gamma$.}

Before concluding this series of simple toy models let us introduce a
further element that may be relevant for the interpretation of the
observations 
{  (see Sect.\ref{sec:template}) 
and that is physically motivated, as we discuss in 
Sect.\ref{sec:discussion}}. We generate
again a nonaxisymmetric system but now this is dominated by circular
velocities in its inner coronas and by radial velocities in the outer
coronas (see Fig.\ref{toymodels_rotrad}(a)). As in the previous case,
the orientation of radial velocities is correlated with the 
major axis of the system. In particular, we choose the toy galaxy to be an ellipse
with $\iota=1$ with solid-body circular velocities in its inner region
and radial velocities in its outer region, {  and we fix $\gamma=2$ in
Eq. \ref{corr_vel}}.
We note that (see Fig. \ref{toymodels_rotrad}(b)-(c)-(d)) (i) the
kinematic axis{  corresponding to the velocity field in the}
 outer coronas forms a small angle
$\psi$ with the projected major axis, as in the previous case
(see Fig.\ref{toymodels2}); (ii) the kinematic axes of the inner and
outer regions form an angle that depends on $j$, that is, it is not simply
$\approx 90^\circ$ as one would have naively guessed on the basis of
Eq.\ref{eq:beckman}; and  (iii) the signature of the two different kinds
of the velocity fields, that is, circular and rotational, is, in this
case, clearly recognizable by looking at the orientation of the 
{  kinematic axes defined 
by the velocity field in the 
inner
and outer coronas, respectively}{.  It is interesting to note that for some
values of $j$ this simple model gives{  values of } $\psi$ of { 
  the order of} a few tens of degrees, so that the two kinematic axes
{  (i.e., the inner and the outer one)} are oriented almost in an
anti-parallel way: it is therefore possible, for example by changing the
shape of the ellipse and{  by taking a larger} 
value of the correlation exponent
$\gamma,$ to further reduce this angle.

\begin{figure}
 \includegraphics[width = 3.5in]{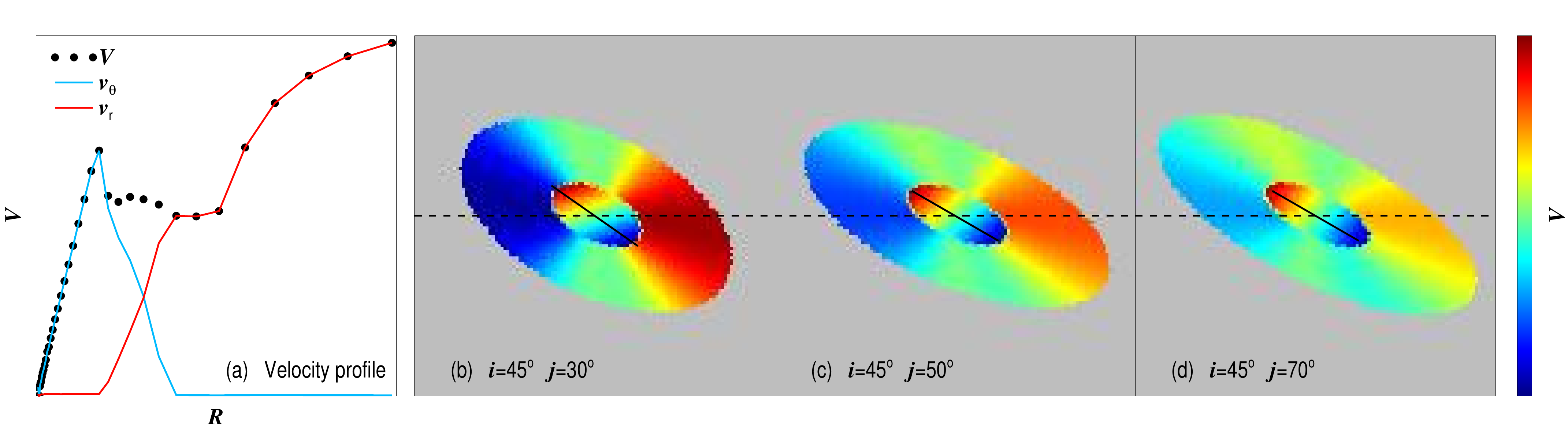}
 \caption{
  {  
 (a) Velocity profile of the toy model with rotational ($v_c$) and radial ($v_r$) velocities
  in the inner and outer coronas, respectively. 
(b)-(c)-(d) Projected velocity field for different values of the angles  $i,j$. 
The kinematic axis in the inner region is shown  as a solid line and 
the kinematic axis in the outer region is shown  as a dashed line.}
}
\label{toymodels_rotrad} 
\end{figure}


\section{Modeling the two-dimensional velocity fields of a galaxy}
\label{sec:template}

In this section we first briefly review the standard methods for
characterizing the observed  two-dimensional velocity field of a galaxy{  under the assumption of axisymmetry}, 
that is, the RDM and the RTRM, and for the detection of noncircular
motions. We then consider the case of a nonaxisymmetric system and we
introduce the REM, discussing its main features.
 {  Finally, we consider how to generalize 
the REM in a physically motivated way. }


\subsection{The rotating disk and tilted-ring models} 
\label{rot_disc_mod}

As mentioned above, observations of galaxy velocity fields have been
interpreted to support, at least to first order, the picture that disk
galaxies are essentially axisymmetric systems in concentric circular
rotation in a plane about a central axis. The velocity of this
rotation, that is, the rotation curve, varies with the radius from the
galactic center, and is assumed to be determined by the radial
distribution of mass within the galaxy.  This situation occurs if
emitters move in an axisymmetric plane and on stable closed orbits,
that is,  on stationary circular orbits. Given this situation, the natural
template {  employed} to fit an observed velocity field {  is a disk} 
with circular velocities such as those measured along the kinematic axis of
the {  projected image of a real} galaxy; in order to find the best RDM one minimizes the
residuals, that is, the difference between real and model velocities, with
respect to the inclination angle $i$. The residuals map traces
noncircular (e.g., radial, random, {nonplanar,} etc.) motions.
{  Therefore, in this case there 
is a single free parameter, the inclination angle,
to be determined by the minimization procedure: others inputs from 
the observations are the systemic velocity and  the angular coordinates of the 
center of the galaxy. In addition,  to further characterize
the orientation of a galaxy,   
the position angle of the
galaxy major axis is usually determined,  measured from north through east: however this 
angle does not necessarily enter into the minimization procedure.}

A refinement of this method is the so-called RTRM, introduced by
\citet{Begeman_1989} for the case of HI observations, but that of
course can be generalized to other kinds of emitters. 
This assumes that{  a} galaxy, again treated as an axisymmetric
system, can be described as a set of concentric rings where each ring
is characterized by a fixed value of the HI surface density, of the
circular velocity $v_c(r),$ and of the orientation angles (i.e., the
inclination angle $i$ and position angle of the galaxy observed major
axis $\phi$\footnote{We adopt the standard convention according to
  which the position angle of the galaxy major axis is measured from
  north through east \citep{Beckman_etal_2004}.}). The three ring
parameters $v_c$, $i,$ and $\phi$ are solved through an iterative
procedure and  appropriate algorithms have been developed to this
aim: for instance, within the GIPSY package (Groningen Image
Processing SYstem;{  see} \cite{van_der_Hulst_etal_1992}) there
is a routine that fits a set of so-called tilted rings to the velocity
field of a galaxy. The code fits a circular model to the velocity
field by adjusting the ring parameters (namely, the kinematic centre,
the inclination, the position angle, and the systemic velocity), so as to
have a list of ring parameters as a function of radius. All the ring
parameters are simultaneously fitted with a general least-squares
fitting routine \footnote{  We note that the GIPSY task ROTCUR has also
  the option of computing radial motions that describes, for a disk, the second
  term in Eq.\ref{eq:beckman}.} .
%

%


\subsection{The radial ellipse model}
\label{exp_ell_mod}

The radial ellipse model (REM) has, by construction, the three main
characteristics of the class of nonaxisymmetric toy models illustrated in Fig.\ref{toymodels2} and with velocities as in
Eq.\ref{corr_vel} (and that are common to the class of objects formed
in the gravitational collapse of isolated self-gravitating cloud of
particles; see discussion in Sect.\ref{sec:discussion}); (i) it hypothesizes a
nonaxisymmetric system where (ii) its velocity field in the outermost
regions is dominated by radial motions such that (iii) radial
velocities are correlated with the system major axis.  The radial
velocities are taken to be directed outwards  {  (as noted above, the
observed velocity field is symmetrical with respect{  to} a change of sign
and thus we could take radial inwards velocity as well and
consider a rotation of 180$^\circ$ of the projected image)}.  The REM
{  must be minimized with respect to} four parameters: the two angles $i,j$ (see above
for definitions), the flatness parameter of the ellipse $\iota$
(Eq.\ref{iota}), and the correlation exponent $\gamma$
(Eq.\ref{corr_vel}).  
{  In addition, as for the RDM, other inputs that must be determined 
from the observations
are the systemic velocity and the angular coordinates of the center of the galaxy.}

In this {  situation} the residuals between the model and
the observed velocity fields trace nonradial (e.g., circular, random,
etc.) motions.  Of course the REM is just a very rough template of the
class of objects with the three characteristics discussed above:
however it is sufficiently simple and versatile to allow a reasonably
good fit of both simple toy models with \textit{a priori} assigned properties{  and observed galaxy velocity fields}.  
For what concerns  {  the latter}, our
{primary goal} is to {illustrate} that a simple template, with a
completely different velocity field than a rotating disk, is able to
fit the data as well as the standard  {   RDM} that  {  is} usually
employed.

{  The physical motivation for introducing the
  REM is that it captures the kinematical properties of a simple class
  of objects formed through the gravitational collapse of an isolated
  cloud of particles that initially breaks spherical symmetry. However
  these systems  are not only dominated by radial motions in their
  outer parts but they are also generally characterized by having an
  inner region where instead rotational motions are predominant 
  (see discussion in Sect.\ref{sec:discussion}). 
  In this situation it is possible to introduce a simple refinement of
  the REM, namely a combination of the REM and of the RDM.

  Namely, in
  the outer parts of the system the velocity field is fitted by a REM
  while  the inner parts, closer to axisymmetry, are fitted by a RDM. It is
  therefore possible to take into account, in a very simple way,  the
  heterogeneous nature of the velocity field of this class of systems. 
  The toy model
  illustrated in Fig. \ref{toymodels2} encodes
    the properties of the
  joint model RDM+REM.  
 The manner in which the motions change from being circular-dominated
 to radial-dominated as a function of the distance from 
 the center of the galaxy can then be optimized for each galaxy and this is 
 of course quite complicated to do for a general case. Below we consider 
   an illustrative example of a THINGS galaxy
  where the transition between circular and radial motions is 
  assumed to be a simple step function.}

\section{Two-dimensional velocity fields of THINGS galaxies}
\label{things}

\subsection{The data} 

The THINGS survey is a high spectral and spatial
resolution survey of HI emission{  of}  34 nearby galaxies obtained using
the NRAO Very Large Array \citep{Walter_etal_2008}.  The HI disks are
better suited than optical emission to study galaxy kinematic as
they allow an entire 2D mapping of the velocity field.
In addition, while in the past HI images of galaxies lack angular
resolution, this survey has a much higher resolution, making it a
unique sample for the study of galaxy kinematics. Indeed, in order to
determine for instance local motions within the disks of galaxies
induced by substructures, one needs to resolve the size scales
associated with features that cause these motions, such as bars,
spiral arms, and oval distortions.

In order to measure high-precision rotation curves
\cite{deBlok_etal_2008} considered a sub-sample of 19 galaxies from
which they excluded galaxies with a low inclination (i.e., $i <
40^\circ$) to avoid large uncertainties in the de-projected rotational
velocity, tidally disturbed galaxies, and those galaxies with an
inhomogeneous and anisotropic velocity field.

We note that the inclination angle can{  generally} be measured under the
assumption that a galaxy is a disk. In this way the inclination
angle of a very nonaxisymmetric object may be overestimated: 
{  simply put, a high inclination galaxy for a RDM  corresponds to a
  very prolate galaxy in the framework of a REM.}  For this reason, in
addition to the \cite{deBlok_etal_2008} sub-sample, we have considered
a further 9 galaxies with estimated (under the assumption mentioned above)
inclination angle $i < 40^\circ$. In addition, we have included in the
sample some galaxies (e.g., NGC 3077, NGC 4449, NGC 5194) that have
strong tidal interactions with neighboring galaxies.  As we show
below, all the galaxies that we considered, except NGC 5194 that has a
very peculiar velocity field, show similar properties from the point
of view of the fitting of their 2D velocity field with a RDM
or with a REM.

\subsection{Fitting the two-dimensional velocity field} 

In order to make the fit with a RDM or with a REM, each galaxy was
coarse-grained with a grid of $64\times64$ equally sized cells and the residuals
between the observations and the best fits were computed on such a
grid.  To find the best-fitting RDM or REM, we need to determine the
best fitting parameters, that is, the inclination angle $i$ for the RDM
and $i,j,\gamma,\iota$ for the REM.  We do this by minimizing the
residuals between the model, computed for generic values of the
parameters, and the actual data on each grid cell. To do so we compute
first, for each grid cell, {labeled by $\alpha$} and centered on
projected coordinates $x_\alpha^\prime,y_\alpha^\prime$, the polar
coordinates as defined above (see Eq.\ref{eq:beckman-tran}):
\begin{eqnarray}
r_\alpha &=& \sqrt{(x_\alpha^\prime)^2+(y_\alpha^\prime)^2} \\ \nonumber
\eta_\alpha &=& \arccos(x_\alpha^\prime/r_\alpha) \\ \nonumber
R_\alpha &=& r_\alpha \sqrt{\cos(\eta_\alpha)^2+
\sin(\eta_\alpha)^2/\cos(i)^2} \\ \nonumber
\theta_\alpha &=& \arctan(\tan(\eta_\alpha)/\cos(i)) \;.
\label{unproj-proj}
\end{eqnarray}
Subsequently, for the case of the RDM, we fix the value of the inclination
angle $i$ (the inclination angle was varied between 20$^\circ$
and 70$^\circ$ with $\Delta i=1$) and we use Eq.\ref{eq:beckman} (with
$v_R=0$) to compute the LOS velocity of the rotational model, denoted {  as}
$v_{los,model}^\alpha$.  We note that in the case where the unprojected
size of the galaxy is larger than the maximum distance at which the
LOS velocity profile extends, we perform a linear fit over the last
five points of $v_{los}(R)$ and then extrapolate using this fit to a
higher radius.  Finally, in order to get the best-fitting inclination
angle, we minimize the sum of the residuals in all the cells with
respect to $i$: 
\be
\label{residuals}
{\cal{R}}= \sum_{\alpha} |v_{los}^{\alpha} - v_{los,model}^{\alpha}| \;,
\ee
where for the RDM ${\cal{R}}={\cal{R}}(i)$ 
\footnote{  We note that the value of the position angle 
of the galaxy major axis is irrelevant for the minimization 
of Eq.\ref{residuals} and is therefore not reported. 
In addition, in the RDM the center of the galaxy image is computed 
by considering only the inner pixels. 
}.   

For the REM case, the  angle $i$ was  varied in the range between 
20$^\circ$ and 70$^\circ$  and $j$ in the range between 
20$^\circ$  and 70$^\circ$ and between 110$^\circ$ and  160$^\circ$
with a resolution
of $\Delta i= \Delta j=10^\circ$. In addition the flatness parameter
$\iota$ was varied in the range [0.3, 0.9] with $\Delta \iota=0.3$ and
the exponent $\gamma$ in the range [1,6], but constrained to assume
the values 1,2,4,6.  As the model is determined by four parameters,
the numerical accuracy with which we can determine their values must
be lowered.  In particular, given this choice of parameters, each
galaxy is compared with about 1000 templates.  For each set of values
of the parameters we numerically calculate the projection onto the
sky with the same procedure as that used for the analysis of the toy models
discussed in Sect.\ref{sec:fits_sims} and we coarse grain such an
image; we then compute the value of the velocity
$v_{los,model}^{\alpha}$ in each coarse-grained cell and we minimize
again Eq.\ref{residuals}, where in this case
${\cal{R}}={\cal{R}}(i,j,\gamma,\iota)$.

We produced RTRM fits of the sample galaxies with \texttt{kinemetry}
\citep{Krajnovic_2006} to extract the axial ratio, $q$, and the
position angle, $\phi$, as a function of the radius.  For each radius,
\texttt{kinemetry} first produces the Fourier expansion of the
velocity along elliptical coronas. We set the code to make the
calculations for a grid of $11\times11$ coronas with axial ratios
between $q=0.2$ and $q=1.0$ and position angles between $\phi=-90^{\rm
  o}$ and $\phi=90^{\rm o}$. The grid point, for which the
$a_1^2+a_3^2+b_3^2$ sum is minimized\footnote{$a_1$, $a_3$, and $b_3$
  are the ${\rm sin}\,(\phi_\alpha)$, ${\rm sin}\,(3\phi_\alpha)$, and
  ${\rm cos}\,(3\phi_\alpha)$ coefficients of the Fourier expansion,
  respectively.}, is used as a starting point for a Levenberg-Marquardt
least-squares minimization from which a final $q$ and $\phi$ are
obtained. The reason for the choice of the function to be minimized is
that errors in $q$ increase $b_3$ whereas errors in $\phi$ increase
$a_1$, $a_3$, and $b_3$.  The radii for which the fits were done are
the \texttt{kinemetry} default ones and follow the formula
\begin{equation}
 R_n=n+1.1^n \;, 
\end{equation}
where $R$ is the radius expressed in pixels and $n$ are the
non-negative integers.  We made \texttt{kinemetry} to stop fitting
once it reached a radius where velocity data are available for less
than 50\% of the data points.

As an illustrative example in Fig.\ref{Ellradg2_RTRM} we show the
determination of the best-fitting RTRM for the toy model discussed in
Fig. \ref{toymodels2}, namely an ellipse with only radial motions
correlated with the orientation of the major axis. One may note
that the RTRM gives a very good representation of the radial velocity
field with residuals smaller than $20\%$ of the maximum velocity. This
result shows the possible confusion between radial and rotational
motion that may occur in the case of {  nonaxisymmetric systems when} 
radial motions are dominant in the system.
\begin{figure}
\includegraphics[width = 3.5in]{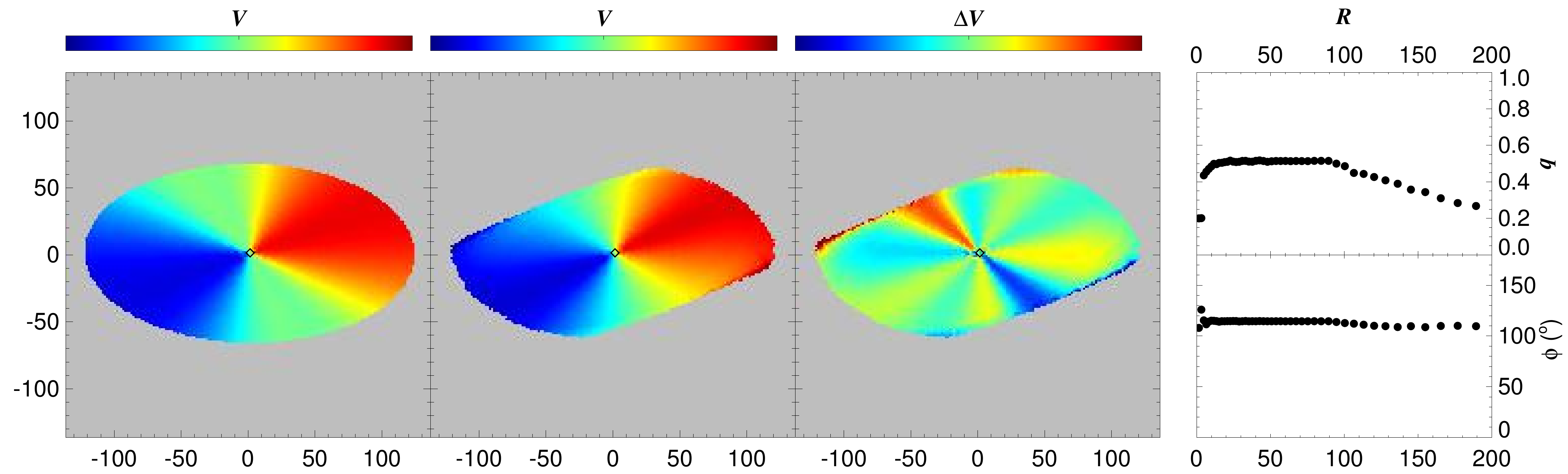}
 \caption{   Best fit RTRM (left panel) and residuals (right panel) for
 the toy model shown in Fig. \ref{toymodels2} for $\gamma=2$, $i=45^\circ$ , and 
 $j=70^\circ$ (see text for details).} 
 \label{Ellradg2_RTRM} 
\end{figure}

\subsection{Results}
In Table \ref{table_THINGS} we summarize {  the results
of the best fits for the THINGS galaxies}.
    The best-fit
inclination angle $i$ for the RDM is reported in column (3); column
(4) reports the amplitude of the residuals defined as
\be
\label{resf} 
f_{res} = \frac{\sigma_{res}}{\sigma_v} \;,
\ee
where $\sigma_{res}$ is the variance of the best-fitting residual
field, and $\sigma_v$ is the variance of the observed velocity field.
We note that both the residual and the velocity field (where, as
mentioned above, we have subtracted the systemic velocity\footnote{The
  systemic velocity for the \cite{deBlok_etal_2008} sample is reported
  in their Table 2, for the other 9 galaxies we have estimated it as the
  velocity of the center. We note that this information enters as an
  overall normalizing factor in the velocity scale.}) have zero
expected mean and $f_{res} < 1$ (in Table \ref{table_THINGS} this is
reported as a percentage).

To quantify noncircular motions, we have computed the cumulative
distribution function of the residuals and have adopted the value
of 95\%\ of the distribution of the residual velocities (in
absolute value) as a representative value of the overall noncircular
($v_{95}^{RDM}$ for the RDM) motion.  
Similarly, we report
the best-fit parameters of the REM $(i,j,\iota,\gamma)$,
the amplitude of the residuals defined as in Eq.\ref{resf} but using
the REM best-fit template, and the value of 95\%
of the distribution of the residual velocities (in absolute value) as
a representative value of the overall nonradial ($v_{95}^{REM}$ for
the REM) motions. 
Finally  we report respectively
$v_{95}^{RTRM}$ and $f_{res}^{RTRM}$ for the RTRM.

\begin{table*}
\begin{center}
\begin{tabular}{|c|c|c|c|c|c|c|c|c|}
\hline
& & & & & &  & &\\
Name & $i$ (RDM)& $f_{res}^{RDM}$& $\left(i,j,\iota,\gamma \right)$ & $f_{res}^{REM}$ & $v_{95}^{RDM}$ &$v_{95}^{REM} $ & $v_{95}^{RTRM} $ &  $f_{res}^{RTRM}$ \\
(1) & (2) & (3) & (4) & (5) & (6) & (7) & (8) &(9)  \\
& & & & & &  & & \\
\hline
{DDO 154* }     & $64^\circ$ & 20\% & $\left(20^\circ,30^\circ,0.3,1\right)$    & 24\% & 13 & 13    &  8   &   13\%    \\
NGC 628          & $20^\circ$ & 54\% & $\left(40^\circ, 20^\circ,0.9,1\right)$ & 67\% & 34 & 30    & 17  &    27\%   \\
{NGC 925* }     & $64^\circ$ & 25\% & $\left(30^\circ, 10^\circ,0.3,1\right)$  & 25\% & 35 & 30    &  24 &    21\%   \\
{NGC 2366* }   & $54^\circ$ & 37\% & $\left(20^\circ, 10^\circ,0.9,2\right)$  & 48\% & 25 & 23    &  17 &    24\%   \\
{NGC 2403*}    & $64^\circ$ & 10\% & $\left(20^\circ, 20^\circ,0.3,1\right)$  & 14\% & 12 & 22    &  11 &      7\%   \\
{NGC 2841*}    & $70^\circ$ & 18\% & $\left(30^\circ, 20^\circ,0.6,4\right)$  & 20\% & 55 & 65    &  30 &      8\%   \\
{NGC 2903*}    & $63^\circ$ & 12\% & $\left(30^\circ, 20^\circ,0.3,1\right)$ & 13\% & 33. & 43    &  21 &      9\%   \\
{NGC 2976*}    & $41^\circ$ & 37\% & $\left(40^\circ, 10^\circ,0.3,2\right)$ & 37\% & 24 & 26     &  24 &    30\%   \\
{NGC 3031*}    & $47^\circ$ & 30\% & $\left(50^\circ, 10^\circ,0.6,2\right)$ & 30\% & 73 & 82     &  72 &    23\%   \\
NGC 3077        & $20^\circ$ & 38\% & $\left(30^\circ, 10^\circ,0.3,1\right)$ & 48\%& 30 & 42     &  24 &    26\%   \\
NGC 3184        &$23^\circ$ & 22\% & $\left(50^\circ, 10^\circ,0.6,1\right)$ & 25\% & 17 & 20     &  12 &    13\%   \\
{NGC 3198*}    &$70^\circ$ & 22\% & $\left(20^\circ, 30^\circ,0.3,2\right)$ & 17\%  & 26 & 36      &  22 &    12\%        \\
NGC 3351        & $39^\circ$ & 13\% & $\left(40^\circ, 10^\circ,0.3,2\right)$& 14\% & 25 & 24      &  10 &      6\% \\
{NGC 3521*}    &$68^\circ$ & 15\% & $\left(40^\circ, 10^\circ,0.3,2\right)$ & 18\% & 44 & 55       &  34 &    12\%  \\
{NGC 3621*}    &$66^\circ$ & 26\% & $\left(20^\circ, 30^\circ,0.3,1\right)$ & 26\% & 51 & 47       &  20 &     12\%       \\
{NGC 3627*}    &$53^\circ$ & 67\% & $\left(20^\circ, 10^\circ,0.6,2\right)$ & 88\% & 123 & 113    &  45 &    24\%         \\
NGC 4214        & $20^\circ$ & 61\% & $\left(10^\circ, 20^\circ,0.9,1\right)$& 63\% & 27 & 19       & 13  &    27\%      \\                    
NGC 4449        &$65^\circ$ & 48\% & $\left(30^\circ, 10^\circ,0.3,1\right)$ & 47\% & 47 & 43       &  21 &    21\%         \\
{NGC 4736*}    &$37^\circ$ & 19\% & $\left(30^\circ, 20^\circ,0.9,1\right)$ & 26\% & 27 & 34        &  15 &     13\%       \\
{NGC 4826*}    &$58^\circ$ & 18\% & $\left(20^\circ, 20^\circ,0.3,1\right)$ & 19\% & 31 & 38        &  30 &     18\%        \\
{NGC 5055*}    &$52^\circ$ & 29\% & $\left(30^\circ, 20^\circ,0.3,1\right)$ & 22\% & 60 & 44        &  15 &      6\%      \\
NGC 5194*      & $20^\circ$ & 92\% & $\left(60^\circ, 40^\circ,0.3,1\right)$ & 100\% & 160 & 100&   58 &   50\% \\
NGC 5236*      & $51^\circ$ & 33\% & $\left(40^\circ, 10^\circ,0.3,1\right)$ & 40\% & 50 & 60      &   23 &   16\% \\
NGC 5457*      & $41^\circ$ &44\% & $\left(60^\circ, 10^\circ,0.3,4\right)$ & 43\% & 43 & 52       &   25 &   22\% \\
{NGC 6946*}   &$32^\circ$ & 22\% & $\left(60^\circ, 10^\circ,0.9,2\right)$ & 27\% & 26 & 39        &   20 &  17\%\\
{NGC 7331*}   &$70^\circ$ & 20\% & $\left(30^\circ, 10^\circ,0.6,2\right)$ & 19\% & 76 & 74        &   50 &  18\%\\
{NGC 7793*}   &$20^\circ$ & 26\% & $\left(30^\circ, 10^\circ,0.9,1\right)$ & 24\% & 26 & 21        &   10 &  10\% \\
{IC 2574}         &$49^\circ$ & 18\% & $\left(60^\circ, 10^\circ,0.6,2\right)$ & 18\% & 11 & 19       &   10  &   10\%  \\
\hline
\end{tabular}
\end{center}
\caption{Parameters and characteristics of the sample of THINGS
galaxies considered in the analysis: (1) Name (the asterisk marks
the galaxies present in the \cite{deBlok_etal_2008} sample) ; 
(2) the best fit inclination angle
obtained by the RDM minimization; 
(3) the amplitude of the residual
fields obtained by the RDM minimization; 
(4) the best-fit parameters
obtained by the REM minimization; 
(5) the amplitude of the residual
fields obtained by the REM minimization;
(6) value (in km/s) of the 95 per cent of the
distribution of the residual velocities (in absolute value) for the
RDM case ($v_{95}^{RDM}$);
 (7) as in (6) but for the REM case
($v_{95}^{REM}$)
(8) as in (6) but for the RTRM
(9)  the amplitude of the residual
fields obtained by the RTRM minimization.
}
\label{table_THINGS}
\end{table*}

In Figs. \ref{ngc628} and \ref{ngc2403} 
 we show the {velocity maps} for the best fitting
velocity field and the corresponding residuals {maps} obtained with
the RDM and with the REM for two representative examples (see
Figs. \ref{ddo154}-\ref{ngc7793} for all the{  other} galaxies)\footnote{In all figures the 
radius is expressed in pixels.}. 

\begin{figure*}
\begin{center}
\includegraphics[width = 5in]{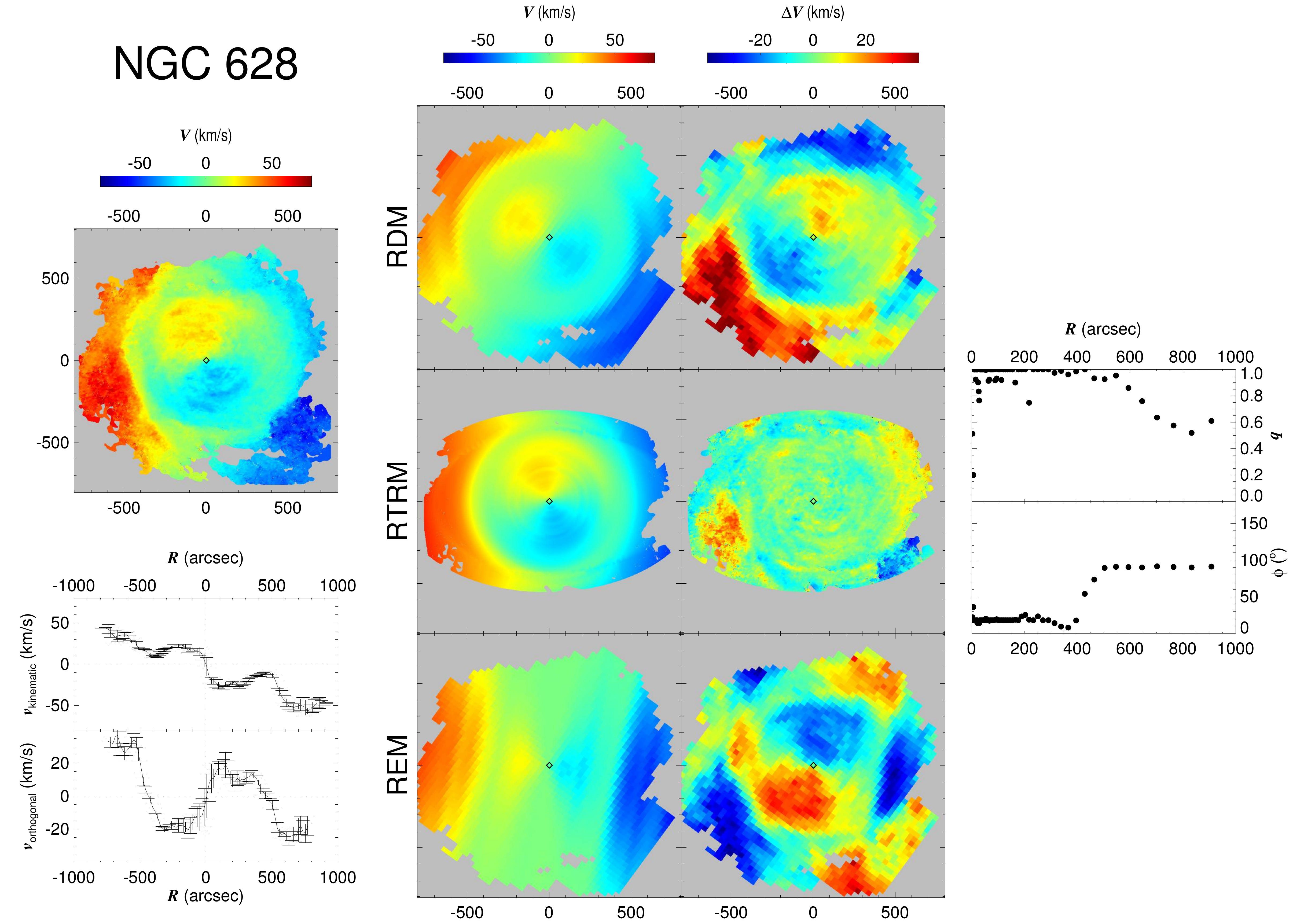}
\caption{  NGC 628 
Left panels: The observed velocity profile (upper panel) and the velocity profile along the kinematic axis and along the axis orthogonal to it (bottom panels).
(The
  color-code of the velocity and residual fields 
  is given in km/s, where the systemic
  velocity of the object has been subtracted.)
Center panels from top to bottom: rotating disk model 
(RDM) velocity field (left), residual 
fields (right). Same as above but for the rotating tilted ring model (RTRM) and 
radial ellipse model (REM).
Right panels: The axial ratio 
(upper panel)  and the position angle
  as a function of the distance from the center (bottom panel).
  }
\label{ngc628} 
\end{center}
\end{figure*}
\begin{figure*}
\begin{center}
\includegraphics[width = 5in]{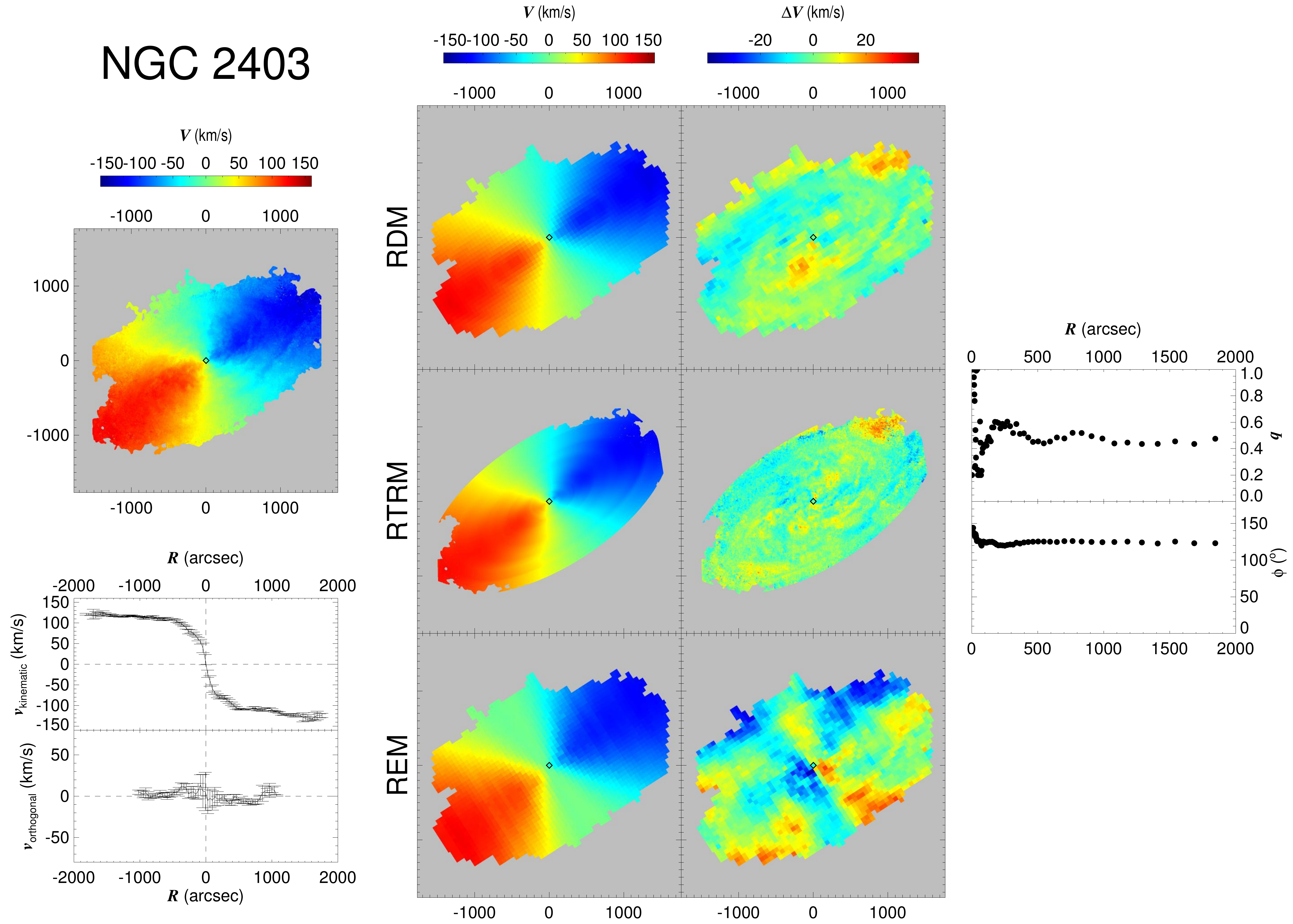}
\caption{  As in Fig. \ref{ngc628} but for NGC 2403.}
\label{ngc2403} 
\end{center}
\end{figure*}

{In addition, }the 1D velocity profile is measured along
two orthogonal slits: one aligned parallel to the kinematic axis and
the {other} orthogonal to this direction (see Figs. \ref{ngc628} and \ref{ngc2403}  for two illustrative examples and 
Figs. \ref{ddo154}-\ref{ngc7793} 
for all the galaxies).

We find that the residuals of the RDM and of the REM are of the same
order of magnitude, that is,  $f_{res}^{RDM} \approx f_{res}^{REM}$.  As
expected, both are larger than the residuals of the RTRM
$f_{res}^{RTRM}$; a similar situation occurs for $v_{95}^{RDM}$,
$v_{95}^{REM}$ and $v_{95}^{RTRM}$.  When the 1D velocity
profile along the kinematic axis is symmetrical, the 1D
velocity profile along the axis orthogonal to the kinematic one has
typically a small amplitude (e.g., DDO 154, NGC 2403, NGC 3351, etc.). In
this situation the residual field has small amplitude and typically
does not present any symmetric (with respect to the center {of the
  galaxy}) patterns. On the other hand, when the residual field has a
large enough amplitude, that is, $f_{res}^{RDM} \approx f_{res}^{REM} >20
\%$ (as for instance, NGC 628, NGC 925, NGC 2366, NGC 2976, etc.)  {then}
the 1D profile along the axis orthogonal to the kinematic
one typically shows a large gradient with localized coherent patterns,
that may correspond to particular structures (e.g., bar) of the
object.

There are two kinds of velocity fields: those
that show coherent patterns in the residuals field of the RDM and
REM best fits, and those that do not.  In the first case (e.g., NGC 2403) the overall kinematic
axis well describes the system at all radii, while in the second case
(e.g., NGC 628) this depends on the distance from the center.  This same
situation corresponds, in the RTRM, to a position angle $\phi$ showing
a radius-dependent behavior.  For example, the case of NGC 628 appears
as a paradigmatic {case} in which the kinematic axis of the inner
region is almost orthogonal to that of the outer
region. Correspondingly the position angle of the RTRM shows a large
variation, almost a step function behavior, as a function of radius.
As discussed above (see Fig.\ref{toymodels_rotrad}), such a situation
is  compatible with the presence of two kinds of velocity
fields, namely rotational in the inner part and radial in the outer
part.  A behavior similar to that shown by NGC 628 is present in
several other galaxies and most notably NGC 3184, NGC 4214, NGC 4826,
NGC 5236, NGC 6946, NGC 7793, and IC 2574.

 {  As we discussed above, in order to take into account the complexity 
 of galaxy velocity fields a single template with radial velocities may not be sufficient.  
 Therefore we have considered, as a simple example and only for illustrative purposes,
a template where at small distances from the center the system is dominated 
by circular motions and at large distances it is dominated by radial motions and is 
 not axisymmetric. Therefore, in practice we have fitted the inner part with a RDM 
 and the outer part with a REM, where the transition between circular 
 and radial motions has been taken to be a step function. Of course 
 it is possible to speculate that a smoother transition would give a better fit
 to the data, but this would clearly be very complicated  implement and goes 
 beyond the scope of the present work. 
  For
   example, in the case of NGC 628 at about 270 arcsec the
   position angle makes a step from a value of about $\phi \approx
   20^\circ$ to $\phi \approx 90^\circ$ (see Fig.\ref{ngc628}). We 
   thus obtain that the inner part is very well fitted by a RDM while the outer part
   can be fitted by a REM with a reasonably good accuracy (see
   Fig.\ref{ngc628b}) although the  RTRM fit still provides a better fit.}
\begin{figure}
\includegraphics[width = 3.0in]{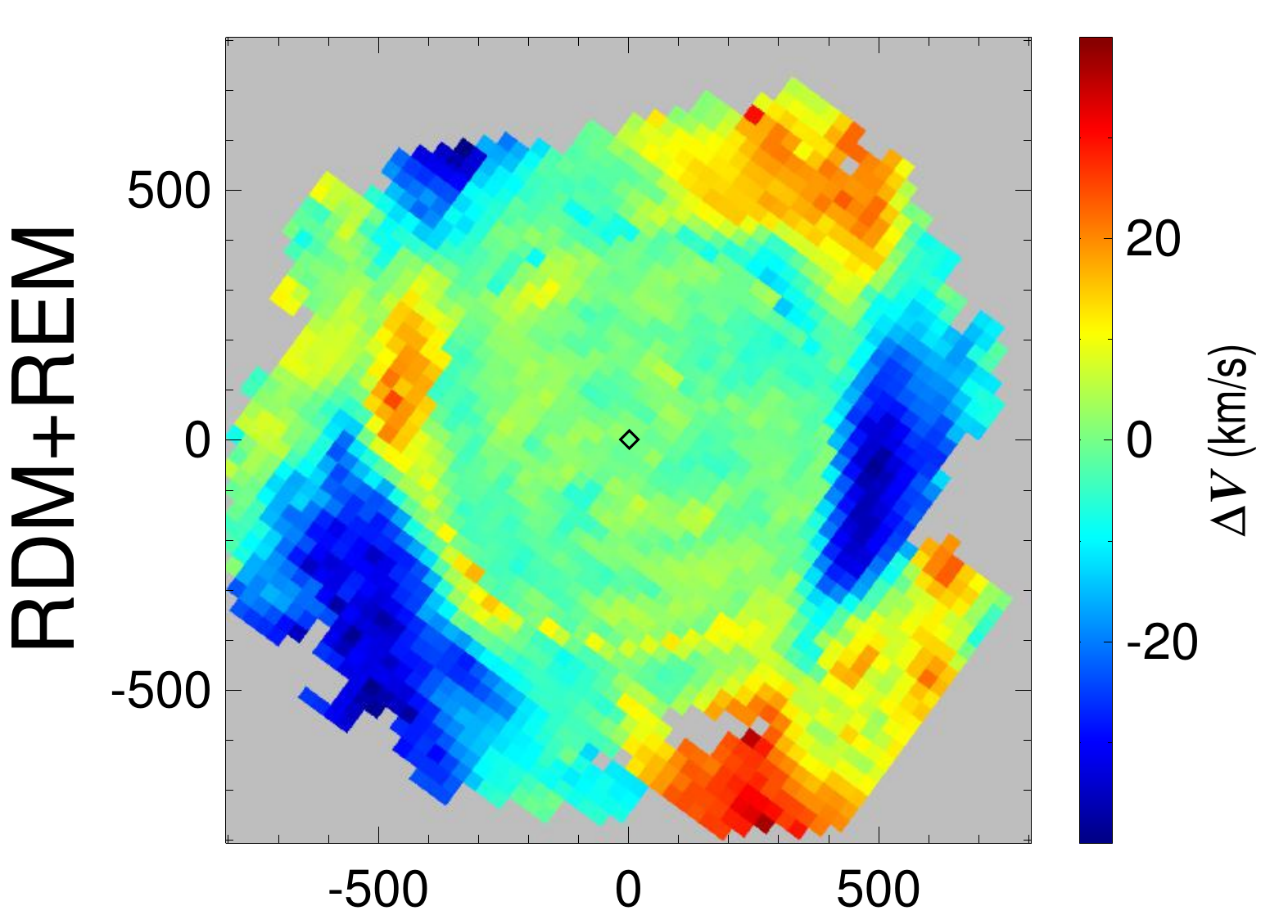}
\caption{   Residuals resulting from 
a fit with a RDM (inner part) and a REM (outer part) to
the galaxy NGC 628 (see text for details).}
\label{ngc628b} 
\end{figure}

{  It is therefore interesting to consider the case of} 
NGC 4826, which seems to show a counter-rotating disk: the
phenomenon of counter rotation \citep[see,
  e.g.,][]{Bertola+Corsini_1999,Khoperskov+Bertin_2017} occurs when
kinematic axes at different radii are oriented in an anti-parallel way.
As discussed above, this situation can arise when there are two
different kinds of velocity fields (i.e., a rotating one in the inner
regions and a radial in the outer ones) in a nonaxisymmetric object
(see Fig.\ref{toymodels_rotrad}) and a modeling with a RDM+REM 
is therefore possible.  Alternatively, it is possible to
consider a situation where there are only radial velocities which are however
oriented in different directions in the inner and outer parts of the
object; for example the inner radial velocity field is directed inwards and
the outer outwards, thus giving rise to patterns in the velocity field that
can be confused with the phenomenon of counter rotation.

\subsection{Discussion}
\label{sec:discussion}
As mentioned above, our aim in this work is not to discuss the
properties of the observed velocity field of each galaxy in detail,
but rather to discuss the method used to determine 
circular and radial motions from the data and to show that one can arrive at very
different conclusions depending on the set of hypotheses on which a
fitting model is based. In particular, we have discussed the role of
the crucial hypotheses, usually adopted, of axisymmetry and circular
motions. These are based on a galaxy model, where stars and other
emitters move in almost circular and stationary orbits in a disk,{  with 
or without a warp},
around the galactic center.

Instead, the main dynamical elements that have inspired the REM are{  characteristic of} a different dynamical model of spiral galaxies
that has emerged from simple simulations of isolated and rotating
overdensities of self-gravitating particles. 
\citep{Benhaiem+Joyce+SylosLabini_2017,Benhaiem+SylosLabini+Joyce_2018}.
      {  In this case}, the combination of an initial rotational
      motion with a strong collapse, which occurs for an
       isolated system driven by its own mean-field gravitational potential, leads very naturally to
      transients, like spiral arms, which have a complex coherent
      spatial distribution and bear a striking qualitative
      resemblance to the real spiral galaxies.  The lifetime of these
      transients can be large compared to the system's characteristic
      time scale $\tau_d \approx 1/\sqrt{G \rho}$, but smaller than
      the collisional time scale $\tau_{coll} $, that  is, these
      transients appear in a range of time $\tau_d \ll t \ll
      \tau_{coll}$ before the system reaches a truly virialized state.
      As discussed in \cite{Benhaiem+Joyce+SylosLabini_2017} the
      physical units can be normalized by fixing the typical velocity
      $\approx 200$ km/s and the typical mass $\approx 10^{11}
      M_{\odot}$: in this situation one obtains a reasonable size for the
      object, namely of the order of tens of kiloparsecs if the collapse process that
      generated the disk and arms occurred much more recently
      (i.e., on a time-scale of the order of 
      1 Gyr) than the formation of the oldest stars in these
      galaxies (with an age $\sim 10$ Gyr).  The precise
      normalization depends however on the properties of the initial
      conditions, as the amplitude of radial velocities and the size
      of the system can greatly vary according to both the amplitude
      of the normalized spin parameter and the shape of the initial
      conditions \citep{Benhaiem+SylosLabini+Joyce_2018}.

In these systems, {   the
origin of the spiral-like arms is related to 
the initial breaking of spherical symmetry and to} 
the nonzero initial
angular momentum.  Indeed, if the evolution leads to a sufficiently
strong contraction of the system during the collapse, some particles
may gain some kinetic energy in the form of a radial velocity oriented
outwards, which adds to the initial rotational velocity.  These are
the particles that are initially placed at the largest distance from
the origin; the dynamical mechanism associated with the monolithic
collapse of the cloud thus amplifies the initial deviations from
spherical symmetry and generates radial motions correlated with the longest axis of the system.
The particles which gain the largest amount of energy will travel,
once faraway from the center of the system, in a gravitational potential
which, to a crude approximation, is spherically symmetric and
stationary; thus, because of approximate conservation of angular
momentum, their radial displacement will be correlated with a decrease
of their angular velocity relative to the center of the structure. As
a result the particles which go furthest will have a smaller angular
velocity than those closer to the center, and will therefore "wind up" less and a
spiral-type structure can result.  Briefly, the outermost particles of
the system that is formed after the collapse are loosely bound, {or are possibly
  even free particles} and have predominantly radial
velocities directed outwards, although they still have a
non-negligible fraction of their velocity in a rotational component.
On the other hand, the system's collapse is generally strong enough to
form an almost virialized and triaxial core with an isotropic velocity
dispersion.  In addition, in an intermediate region between the inner
core and the outer loosely bound particles a{  flat} disk-like configuration
is formed where rotational motions are predominant.
{  Thus, the velocity field that results from these simple systems  
is heterogeneous in nature and strongly scale dependent.}

{  It must be emphasized that the class of models considered by
  \cite{Benhaiem+Joyce+SylosLabini_2017,Benhaiem+SylosLabini+Joyce_2018}
  is very simplistic, not just because of the idealization of the
  initial conditions but also in that it neglects everything but
  gravitational dynamics.  Any detailed quantitative model of a real galaxy 
  will of
  course necessarily need to consider more complex initial conditions
  and also incorporate nongravitational physics, such as  gas dynamics,
  cooling, star formation, and so on.  In this respect we note that in
  standard models of galaxy formation the key element in the formation
  of a disk galaxy is the dissipation associated with
  nongravitational processes.  Instead, in the purely gravitational
  simulations by
  \cite{Benhaiem+Joyce+SylosLabini_2017,Benhaiem+SylosLabini+Joyce_2018}
  disk-like configurations with transient spiral arms are formed by
  purely dissipationless gravitational dynamics if the initial
  conditions break spherical symmetry. For this reason, a more
  complete study of this class of models requires the study of
  hydrodynamical simulations of nonaxisymmetric systems: 
this is   currently an ongoing work 
  \citep{SylosLabini_RCD_DLP_2018}.

For what concerns the problem of cosmological galaxy formation, the monolithic 
collapse of an overdensity may occur in top-down structure formation
scenarios like warm dark-matter models. This process, depending on the
properties of the power-spectrum of density fluctuations, can be
theoretically 
approximated by the gravitational collapse of an isolated cloud
\citep{Peebles_1983} similar to what occurs in the simulations of a
cold collapse.  }

To summarize, the main features of the outermost region of these
systems are: (i) they are not axisymmetric, (ii) they have a radial
velocity field directed outwards that (iii) has a strong correlation
with the direction of the major axis of the system.  These are three
features that are encoded in the REM.  {  In addition}, this class
of system usually presents an extended flattened region which rotates
coherently about a well virialized core of triaxial shape with an
approximately isotropic velocity dispersion.  For this reason the
resulting velocity field has a{  heterogeneous} scale-dependent
behavior and a fitting model that assumes only a single type of
velocity field (rotational or radial) is not suitable to describe the
systems at all radii.




\section{Conclusions}
\label{sec:dicus_concl}

Observed 2D galaxy velocity maps are usually interpreted
under the hypothesis that stars (and other emitters) move in an
axisymmetric disk and in stationary circular orbits around the
galactic center. This hypothesis is corroborated by the observed
velocity gradient in such maps, that is, at least at first order,
consistent with a rotational velocity field.  In order to characterize
galactic kinematics in greater detail, these 2D velocity
maps are fitted to a rotating disk model (RDM) or, {  in order
to take into account the possible presence of a warp in the 
galaxy}, to a rotating
tilted-ring model (RTRM). {  In} both cases, the hypothesis used is that
the galaxy{  is  axisymmetric} and that stars and other
emitters move in circular orbits: {  in the case
of the RTRM it is also assumed that circular orbits at different distances 
from the galaxy's center have
different inclination.}  The residuals, that is, the difference between
the observed and the model velocity fields, are then interpreted to
trace noncircular motions like radial, random, and other motions.

In this paper we have shown that if the system is not assumed to be
axisymmetric then it is possible to elaborate a different
interpretation of the observed 2D galaxy velocity maps.
This is based on a different dynamical model of the outermost regions
of the observed galaxies: namely that stars and other emitters have a
velocity field where radial motions are large and/or 
dominant. Such a model describes the properties of the transient
spiral-like structures that are formed in the collapse of isolated,
initially nonspherical, and rotating clouds of particles in which
is generically formed a rotating disk surrounded by transient
spiral-like arms of which the motion is mainly radial.  The complex
velocity and configurational properties of these structures can be, as
a first rough approximation, described as an ellipse with radial
velocities directed outwards whose amplitude is correlated with the major axis of the ellipse. These properties are encoded in the fitting model
that we have described, the radial ellipse model (REM).

{  We have shown that}, from a numerical point of view, for a
sub-sample of galaxies extracted from the THINGS data, the REM works
as well as the RDM; that is residuals are of the same order of magnitude
both for the case in which the fitting template is a disk with
circular velocities (RDM) and the case of an ellipse with radial velocities
correlated with the major semi-axis (REM).  
{  
It should be noted that the REM is defined by four parameters
  rather than only one, as in the RDM. This is because the former
  describes a system (i.e., an ellipse with a correlation between the
  direction of radial velocities and its major axis) that is much more
  complicated than a simple disk with only circular motions.  
  Thus, from a purely mathematical analysis point of view,
  given that the REM and RDM give similar residuals, 
   if we 
  apply the Akaike information criterion \citep{Akaike}, 
  the REM has a likelihood smaller by a factor of $\sim \exp(-6)$ compared 
  to that for the RDM,
  and therefore is much less favorable. 
  However, although the RDM delivers the best results compared to the
  number of parameters used,  it
  does not take into account the possible 
  complexities of real galaxies, such as
  the fact that axisymmetry is often observed to be broken.
  }

We have also studied the
best fits obtained with the rotating tilted ring model (RTRM) and
found that in this case the residuals are the smallest. However we
have argued that this is possible because such a model allows a
radius-dependent orientation of the kinematic axis and that, in this
way, it may confuse rotational and radial motion, if they are present
at different radii in a given system.
{  Finally we have stressed that, in order to more accurately 
describe a heterogeneous and scale-dependent  velocity field,
a template consisting in a combination of the RDM and the REM
may be more suitable. This joint template represents an alternative 
physically motivated model to the RTRM,  able to 
characterize the velocity fields in nonaxisymmetric systems.
In this respect we stress that 
  the presence of the warp, confirmed in several cases   by observations
  different  from kinematical ones 
    \citep{sancisi_1976,Reshetnikov+Combes_1998,Schwarzkopf+Dettmar_2001,Garcia_Ruiz_etal_2002,Sanchez-Saavedra_etal_2003,Levine_etal_2006,Kalberla_etal_2007,Reyle_etal_2009},  
  is not incompatible with the presence of
  radial motions: that is, the presence of a warp  does not imply, but 
  it is only compatible with, the key 
  assumption of the  RTRM    that stars orbit on circular orbits
  with a different inclination as a function of the distance from the galaxy's center.
}

It is clear that when the mass is weakly bound or even unbound, as
occurs when radial velocities are not negligible, one greatly
overestimates the actual enclosed mass from the dynamical mass
$M^{dyn}(r) \approx v^2(r) \cdot r/G$, that is,  if one assumes stationary
circular orbit. Specifically, if a part of the observed velocity has a
contribution from a radial component, and thus the system has not
reached a truly virialized state, then $M^{dyn}(r)$ gives a greater
estimation of the actual mass: this situation suggests the possibility
that observed velocity curves of external galaxies might require much
less dark matter than what is usually estimated from $M^{dyn}(r)$ if
the outer parts of the galaxy are far from stationary and the motions
are predominantly radial {and spatially correlated in a nonaxisymmetric
  distribution}, rather than rotational
\citep{Benhaiem+Joyce+SylosLabini_2017}.

Concomitantly, as we have shown, in external galaxies it is not
straightforward to disentangle between rotational and radial velocity
fields if the system is not axisymmetric; for our own Galaxy the
measurement of the radial and tangential components of stars'
velocities is now possible with an increasing precision.  In this respect, it
is interesting to mention that, while it has been known for several
decades that the Galactic disk contains large-scale nonaxisymmetric
features, a complete understanding of these asymmetric structures and of
their velocities fields is still lacking.  The recent Gaia DR2 maps
\citep{Katz_etal_2018} have clearly shown that the Milky Way is not an
axisymmetric system at equilibrium, but that it is characterized by
streaming motions in all the three velocity components.  In particular these data 
have confirmed the coherent radial motion in the direction of the
anti-center, earlier detected up to 16 kpc by \cite{MLC_CGF_2016} and
recently extended up to 20 kpc by \cite{MLC_etal_2018}. 

 {  At larger distances, even though the data are very noisy, using
   a statistical method of deconvolution of the parallax errors,
   \cite{MLC_FSL_2018} were able to measure not only significant
   departures of circularity in the mean orbits with radial
   Galactocentric velocities between -20 and +20 km/s but also
   vertical velocities between -10 and +10 km/s, variations of
   rotation speed with position, and asymmetries between
   northern and southern Galactic hemisphere of up to 20 kpc.  Of course 
   the tangential velocity component is still very much larger than the 
   radial one, and this latter can already have an important 
 dynamical effect   \citep{MLC_etal_2018}. For this reason 
 it is crucially important to study the kinematics of our Galaxy 
 in the outermost region of the disk. 
   The next
   GAIA data release, foreseen for 2020, with improved astrometry
   and photometry allowing it to map the velocity field in the outer part of
   the Galactic disk, will be able to map distances larger than $r>
   20$ kpc.}

\begin{acknowledgements}

FSL and DB thank Michael Joyce for collaborations and discussions, and
Roberto Capuzzo Dolcetta for useful comments.  This work was granted
access to the HPC resources of The Institute for scientific Computing
and Simulation financed by Region Ile de France and the project
Equip@Meso (reference ANR-10-EQPX- 29-01) overseen by the French
National Research Agency (ANR) as part of the Investissements d'Avenir
program.  MLC was supported by the grant AYA2015-66506-P of the
Spanish Ministry of Economy and Competitiveness (MINECO).  We
acknowledge the use of the THINGS project data available at {\tt
  http://www.mpia.de/THINGS/Data.html}. We thank Fabian Walter for kindly
giving us access to the data of IC 2574.  We acknowledge the use of
{  the code} \texttt{kinemetry} by \cite{Krajnovic_2006}.  { 
  Finally we thank an anonymous referee for very useful
  comments and suggestions}.
\end{acknowledgements}


\newpage
\clearpage

\begin{figure*}
\begin{center}
\includegraphics[width = 5in]{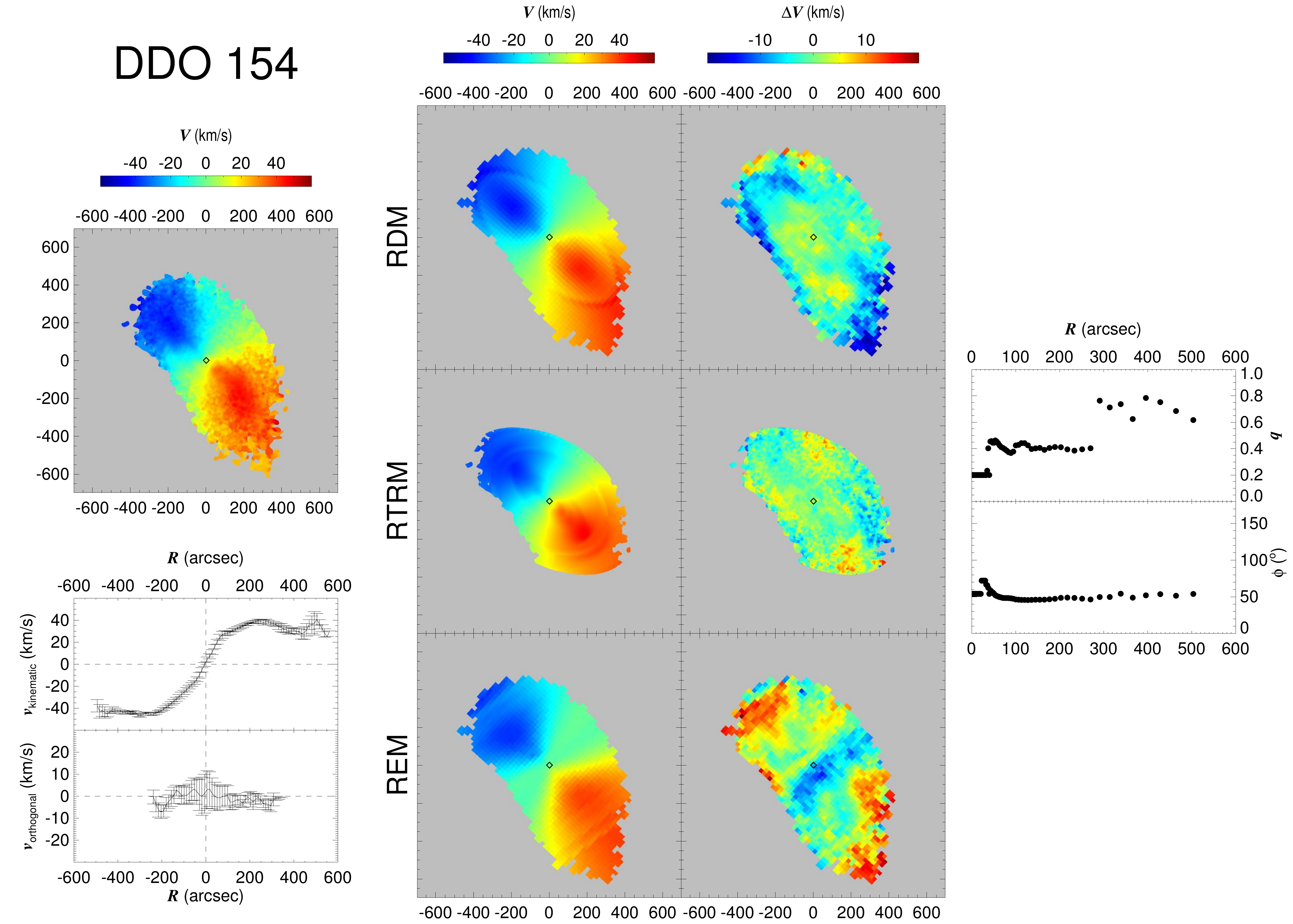}
\caption{As in Fig.\ref{ngc628} but for DDO 154.}
\label{ddo154} 
\end{center}
\end{figure*}

\begin{figure*}
\begin{center}
\includegraphics[width = 5in]{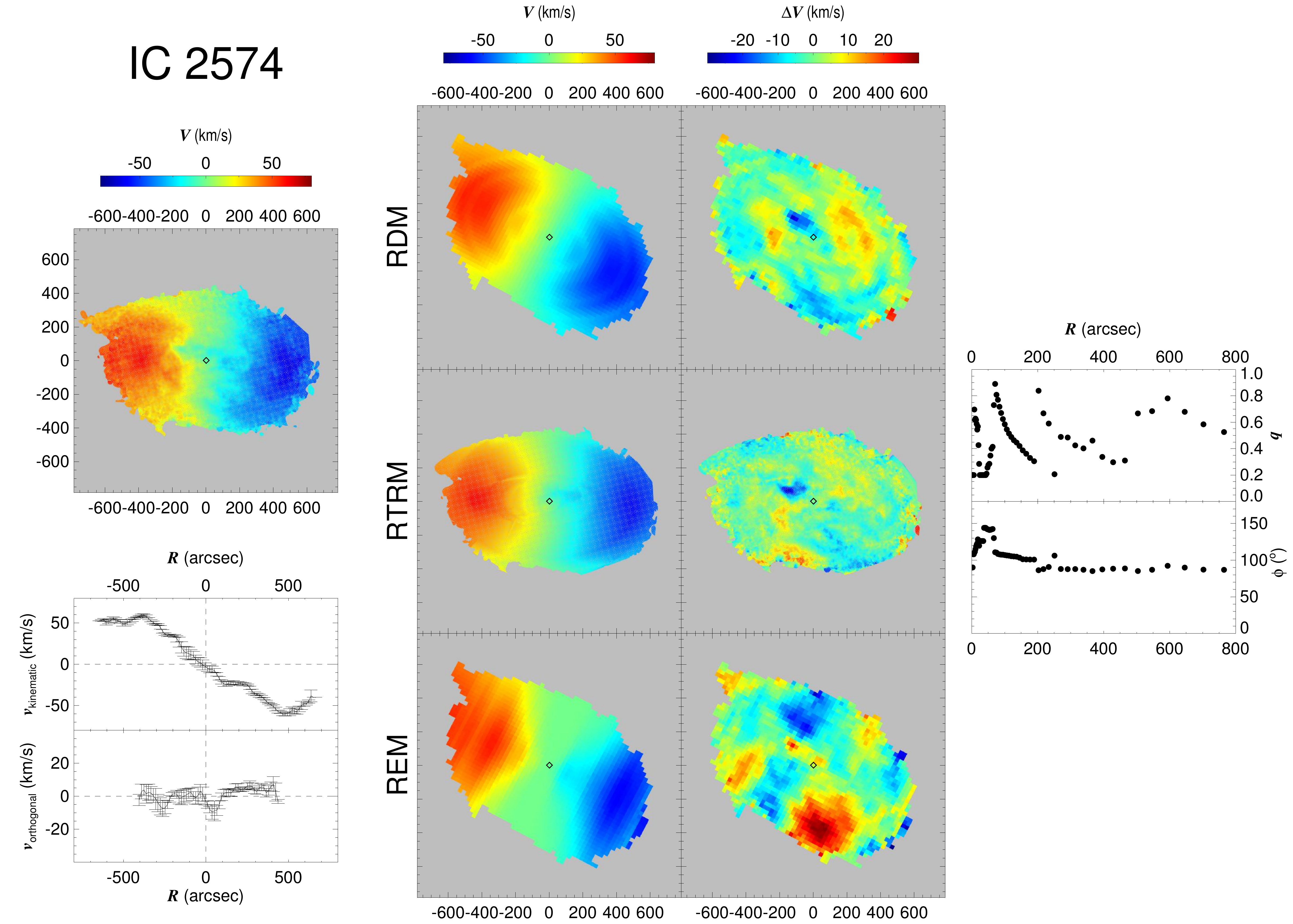}
\caption{As in Fig.\ref{ngc628} but for IC 2574}
\label{ic2574} 
\end{center}
\end{figure*}

\newpage
\clearpage

\begin{figure*}
\begin{center}
\includegraphics[width = 5in]{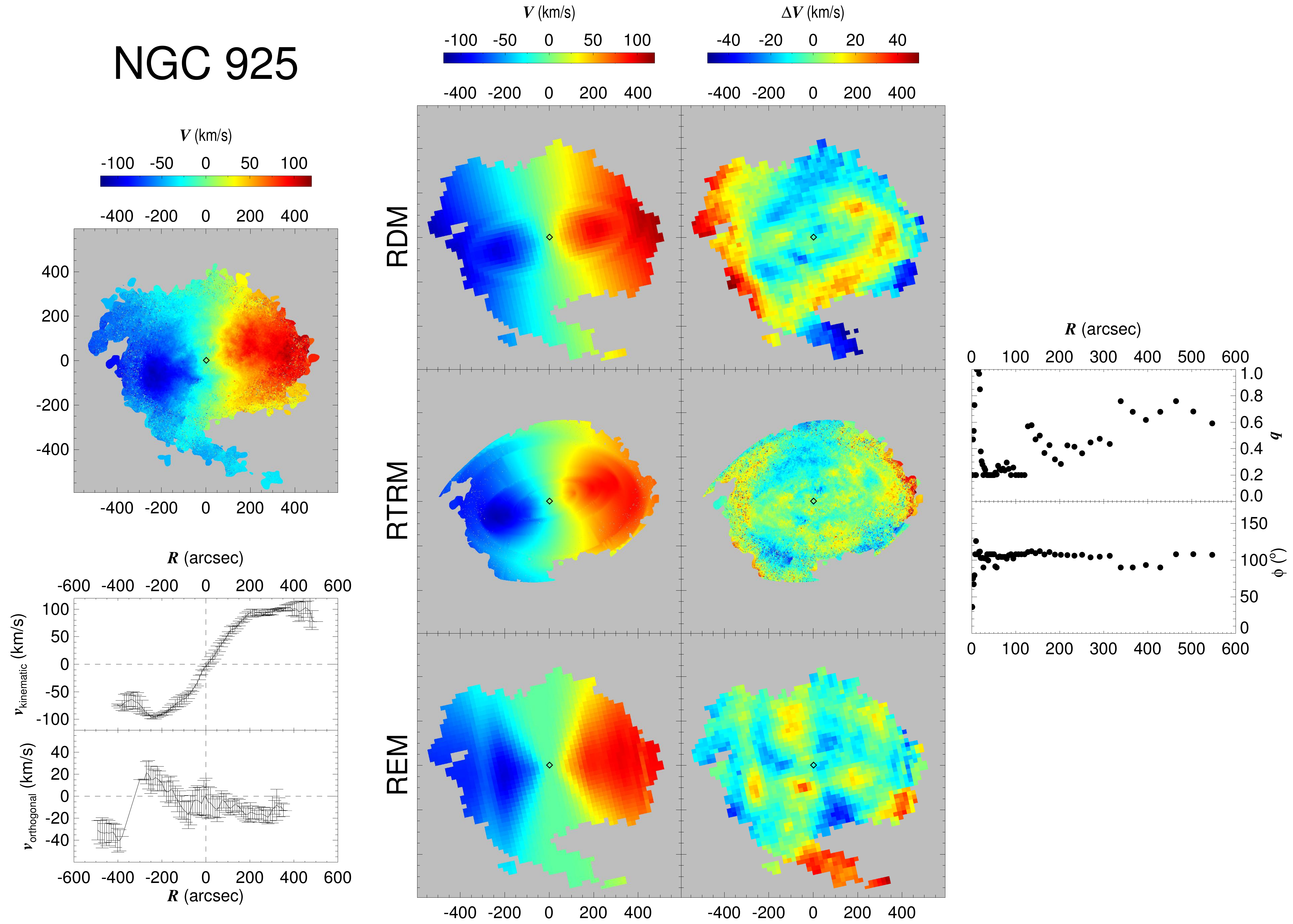}
\caption{As in Fig.\ref{ngc628} but for NGC 925}
\label{ngc925} 
\end{center}
\end{figure*}

\begin{figure*}
\begin{center}
\includegraphics[width = 5in]{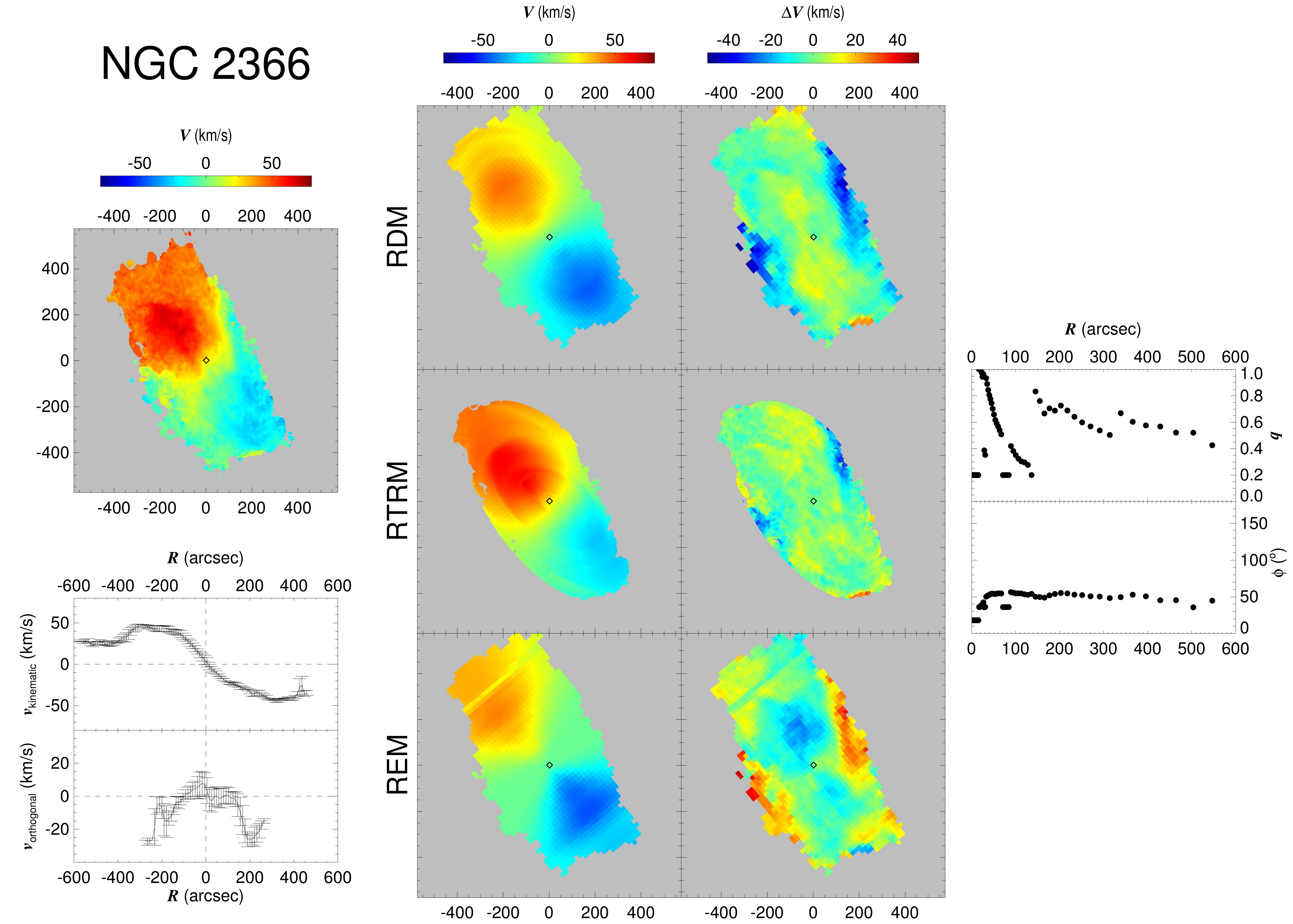}
\caption{SAs in Fig.\ref{ngc628} but for NGC 2366}
\label{ngc2366} 
\end{center}
\end{figure*}

\newpage
\clearpage

\begin{figure*}
\begin{center}
\includegraphics[width = 5in]{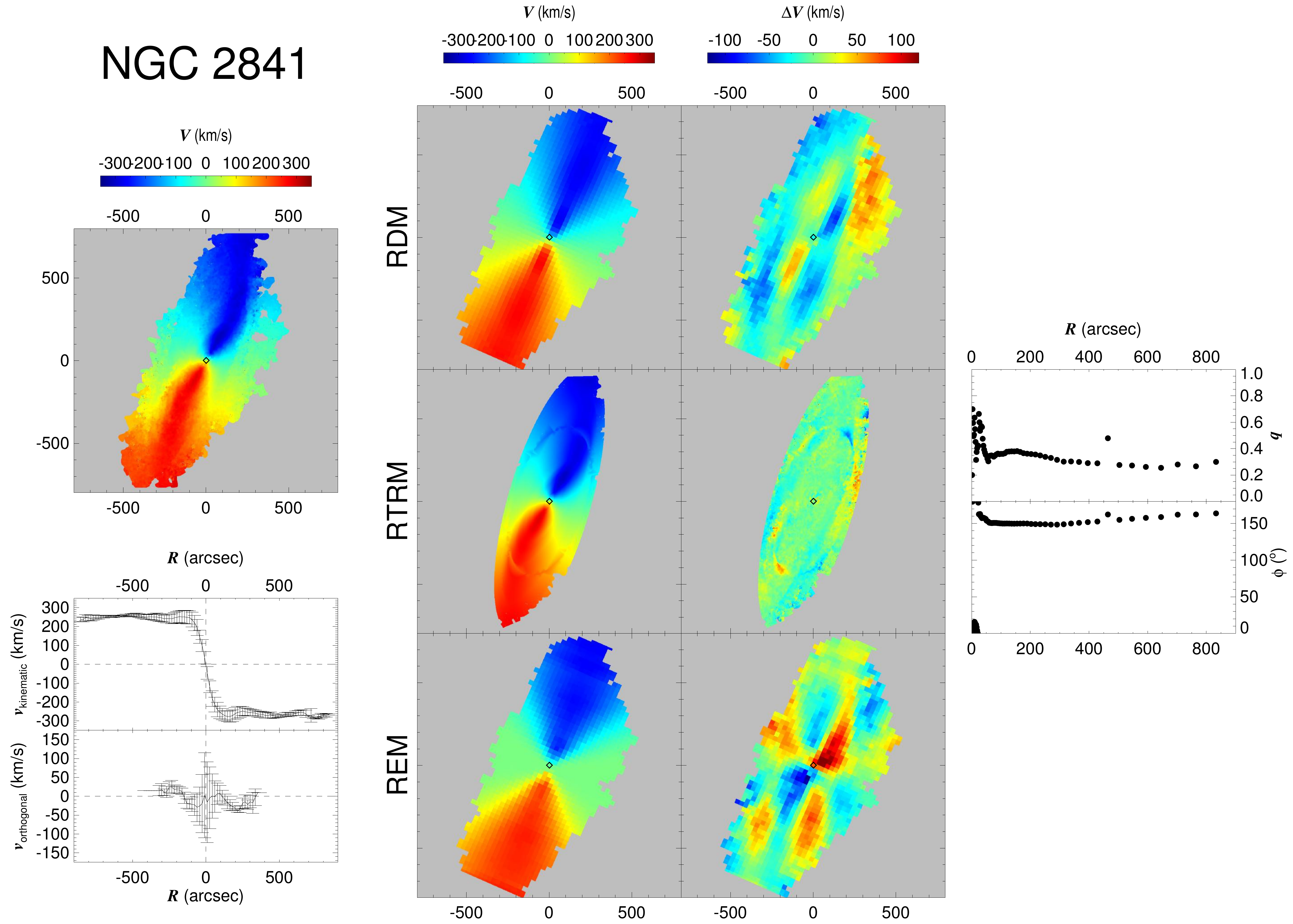}
\caption{As in Fig.\ref{ngc628} but for NGC 2841}
\label{ngc2841} 
\end{center}
\end{figure*}

\begin{figure*}
\begin{center}
\includegraphics[width = 5in]{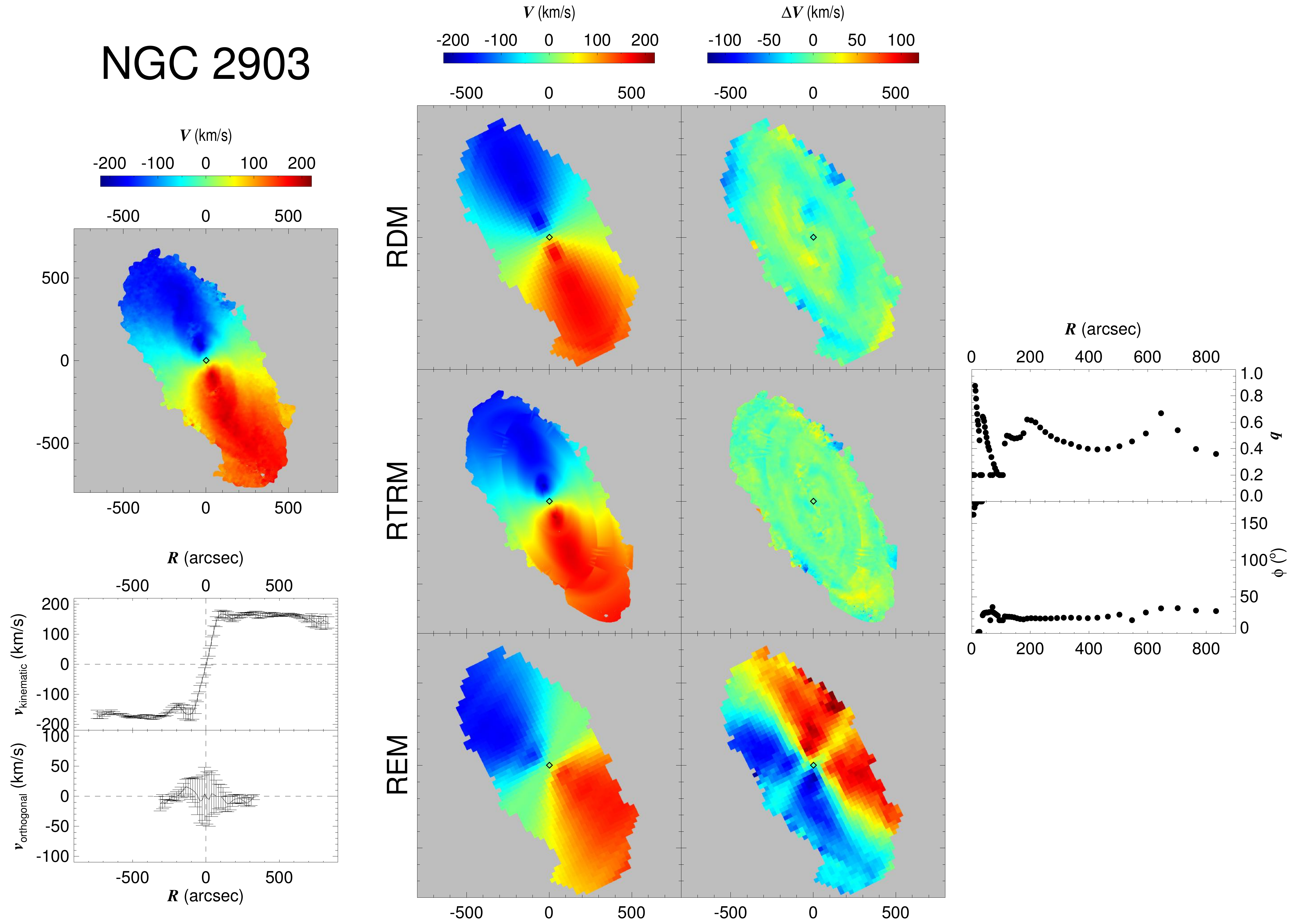}
\caption{As in Fig.\ref{ngc628} but for NGC 2903}
\label{ngc2903} 
\end{center}
\end{figure*}

\newpage
\clearpage

\begin{figure*}
\begin{center}
\includegraphics[width = 5in]{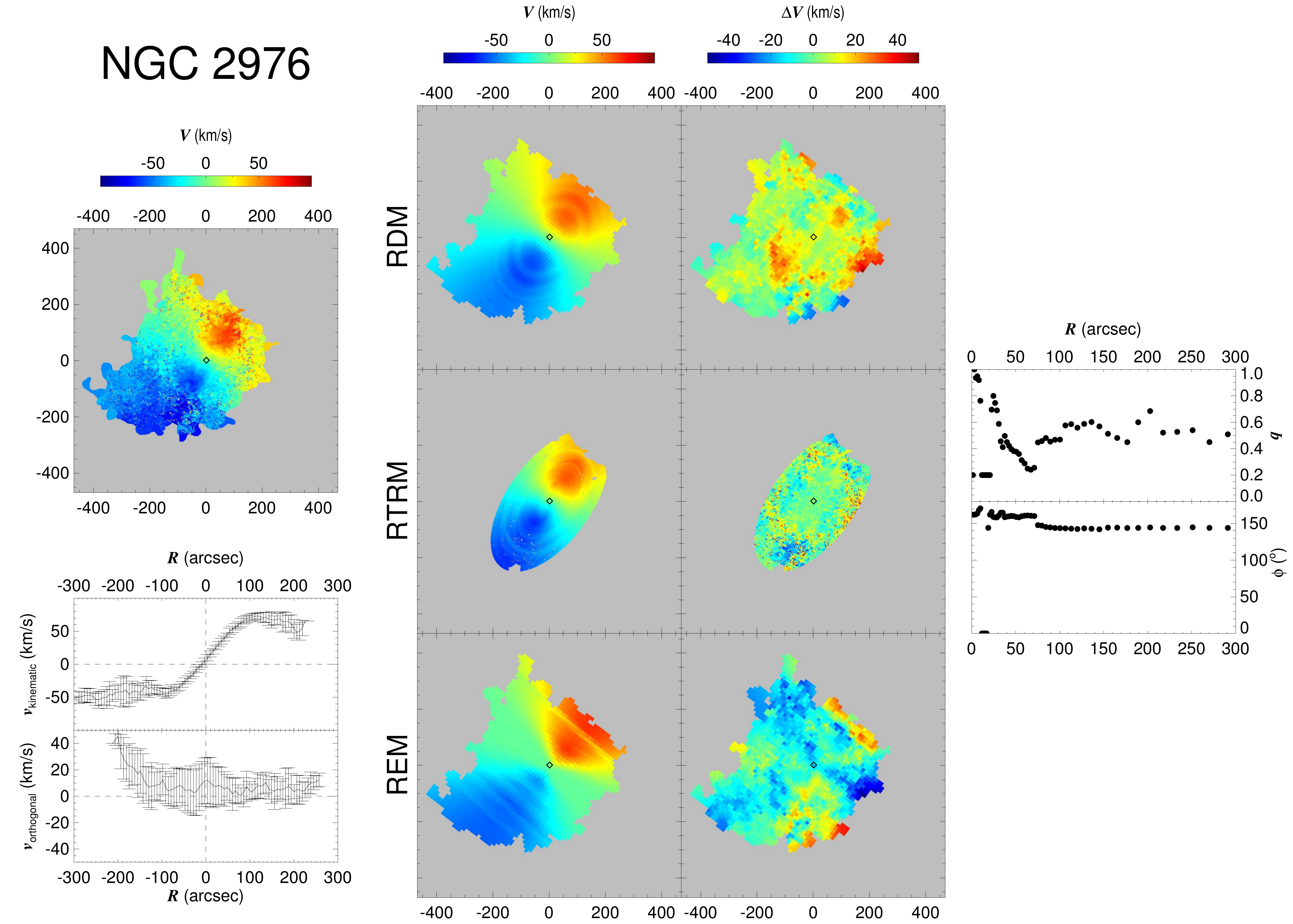}
\caption{As in Fig.\ref{ngc628} but for  NGC 2976}
\label{ngc2976} 
\end{center}
\end{figure*}

\begin{figure*}
\begin{center}
\includegraphics[width = 5in]{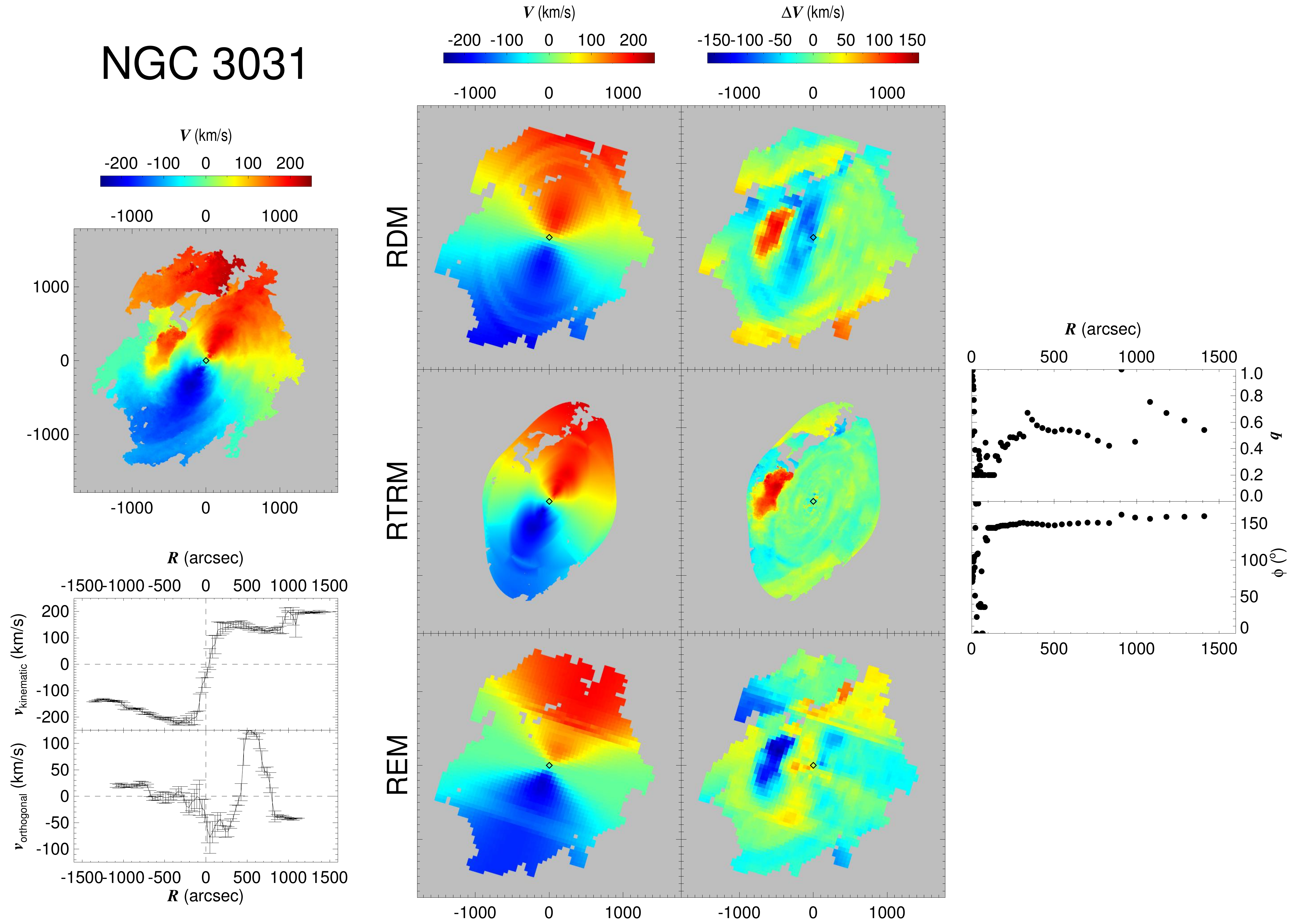}
\caption{As in Fig.\ref{ngc628} but for NGC 3031}
\label{ngc3031} 
\end{center}
\end{figure*}

\newpage
\clearpage

\begin{figure*}
\begin{center}
\includegraphics[width = 5in]{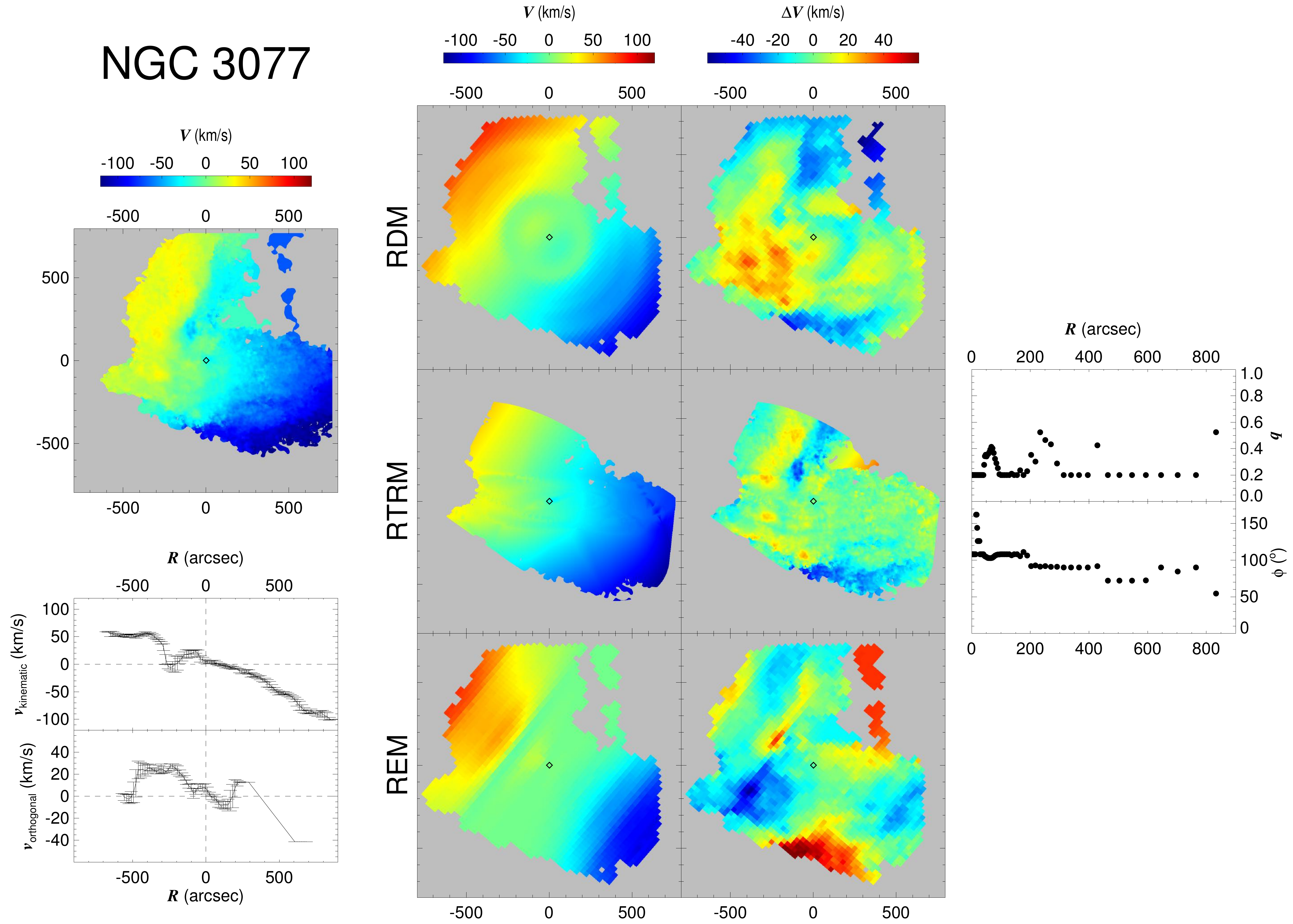}
\caption{SAs in Fig.\ref{ngc628} but for NGC 3077}
\label{ngc3077} 
\end{center}
\end{figure*}

\begin{figure*}
\begin{center}
\includegraphics[width = 5in]{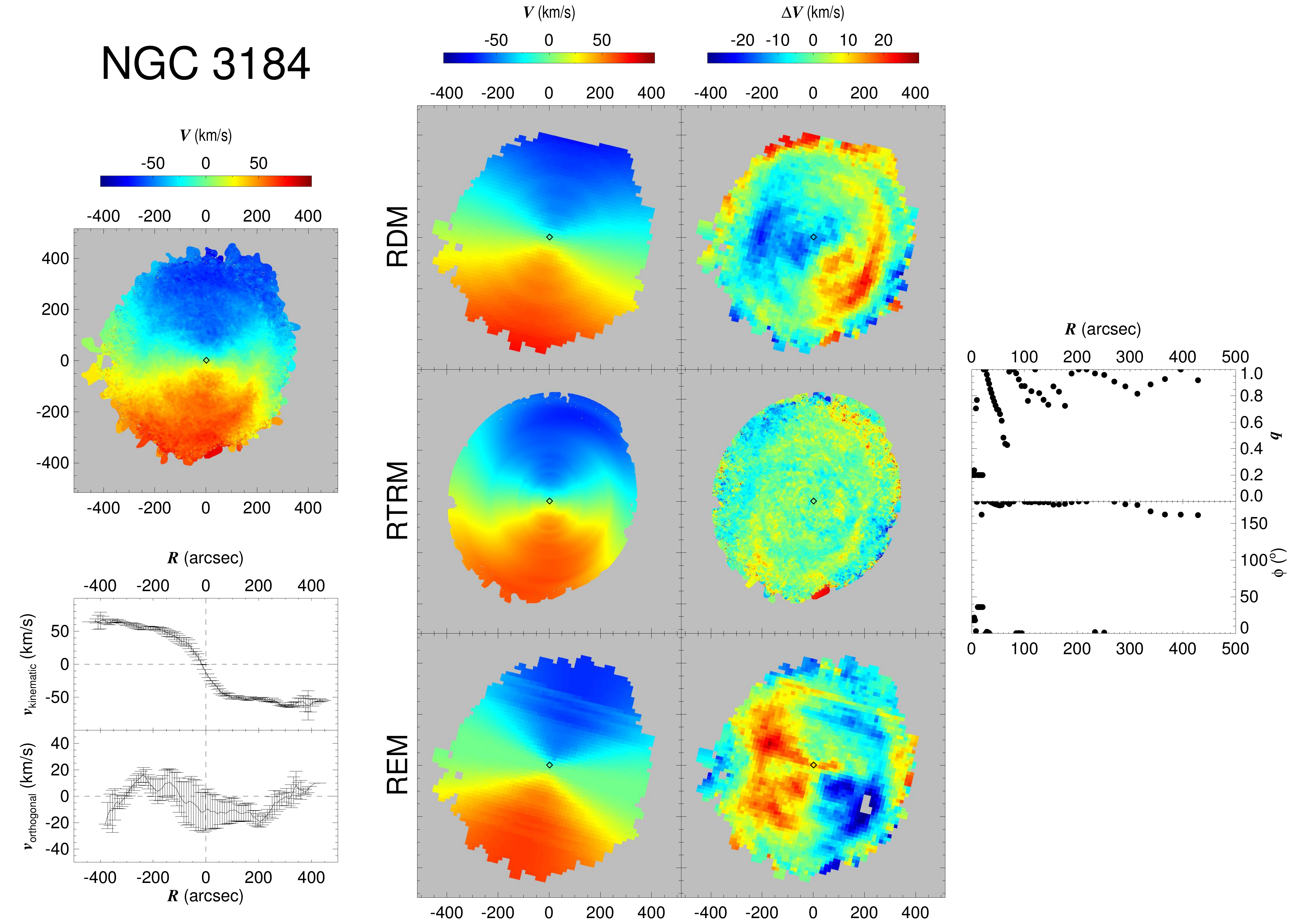}
\caption{As in Fig.\ref{ngc628} but for NGC 3184}
\label{ngc3184} 
\end{center}
\end{figure*}

\newpage
\clearpage

\begin{figure*}
\begin{center}
\includegraphics[width = 5in]{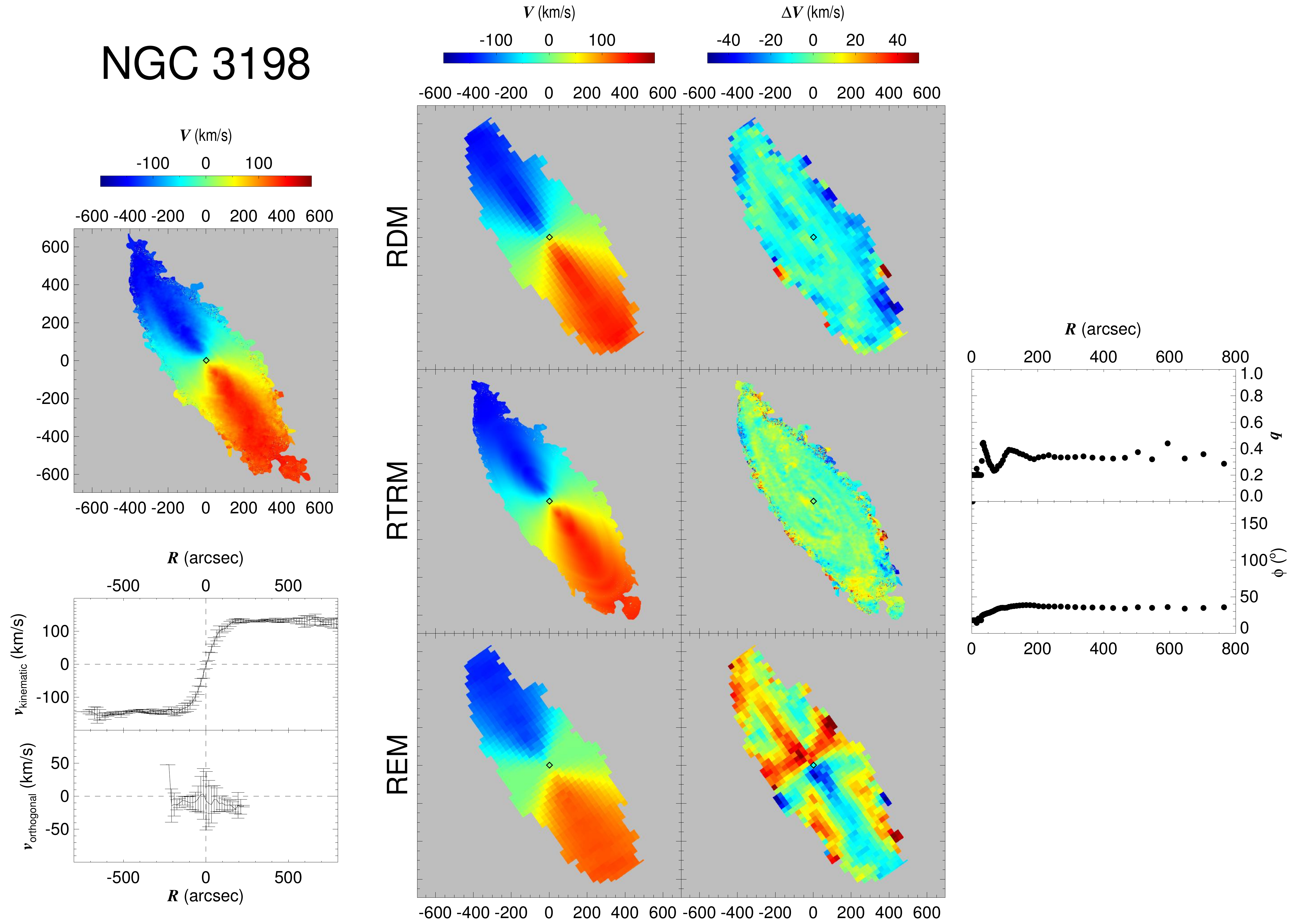}
\caption{As in Fig.\ref{ngc628} but for NGC 3198}
\label{ngc3198} 
\end{center}
\end{figure*}

\begin{figure*}
\begin{center}
\includegraphics[width = 5in]{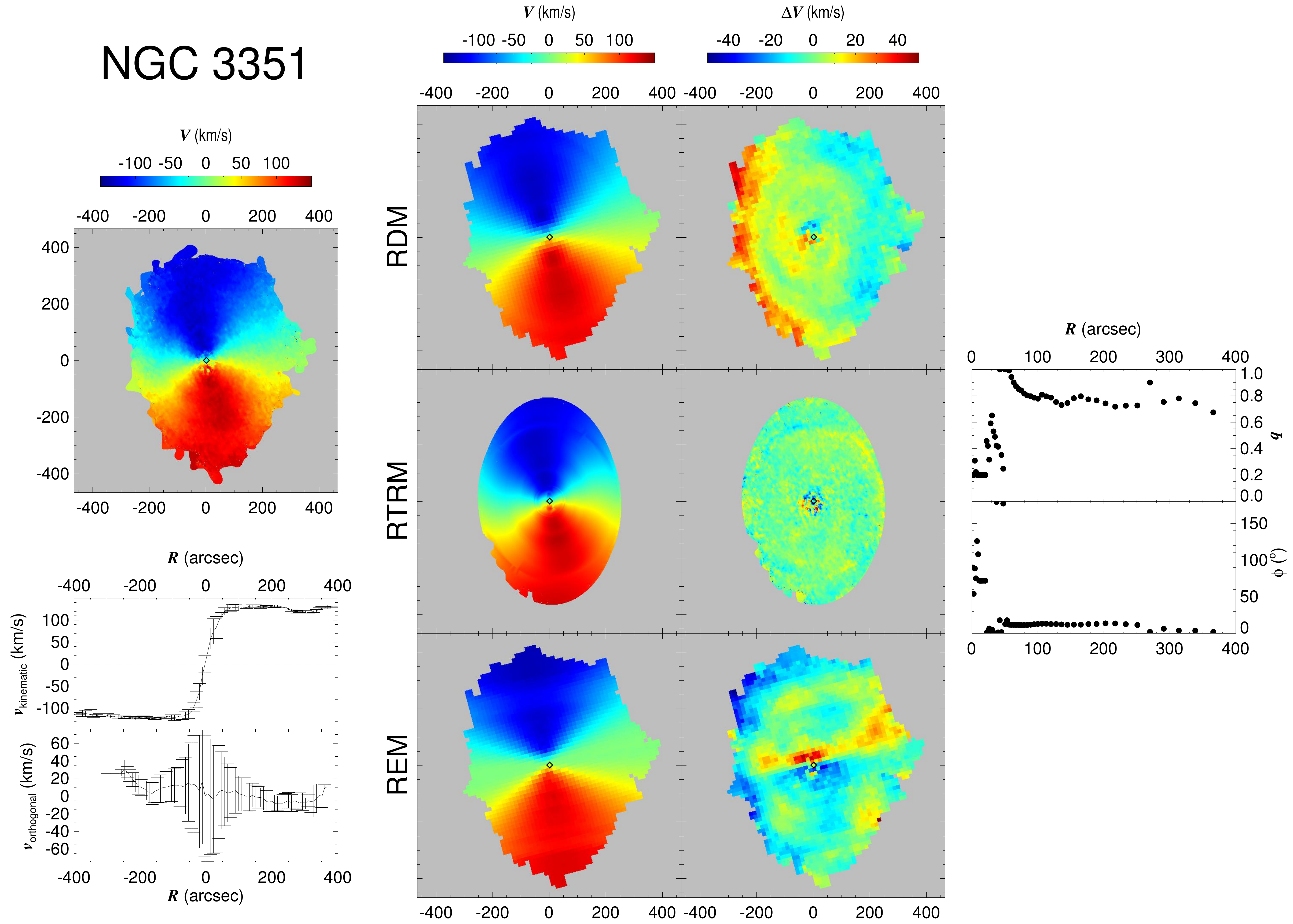}
\caption{As in Fig.\ref{ngc628} but for NGC 3351}
\label{ngc3351} 
\end{center}
\end{figure*}

\newpage
\clearpage

\begin{figure*}
\begin{center}
\includegraphics[width = 5in]{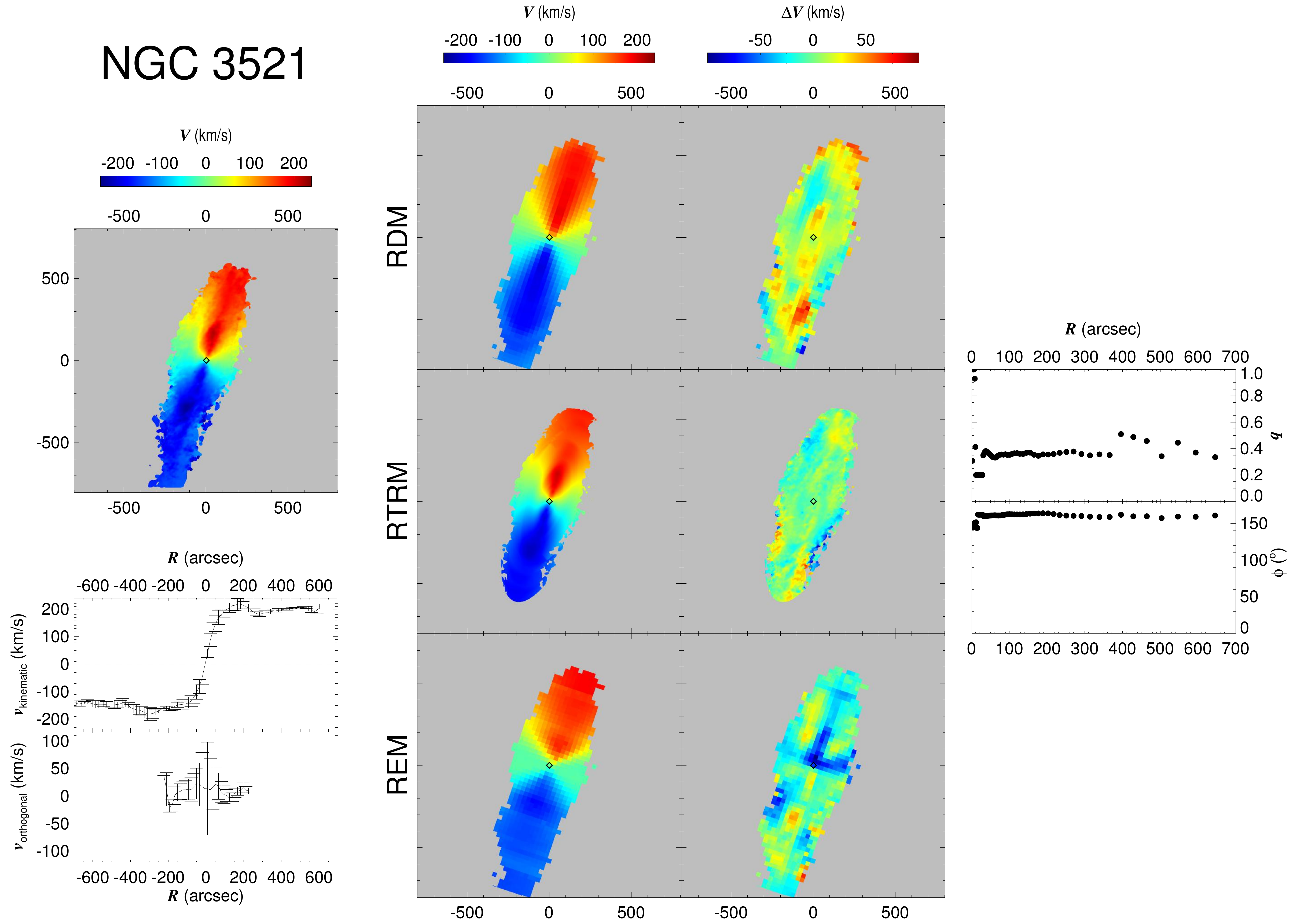}
\caption{As in Fig.\ref{ngc628} but for  NGC 3521}
\label{ngc3521} 
\end{center}
\end{figure*}

\begin{figure*}
\begin{center}
\includegraphics[width = 5in]{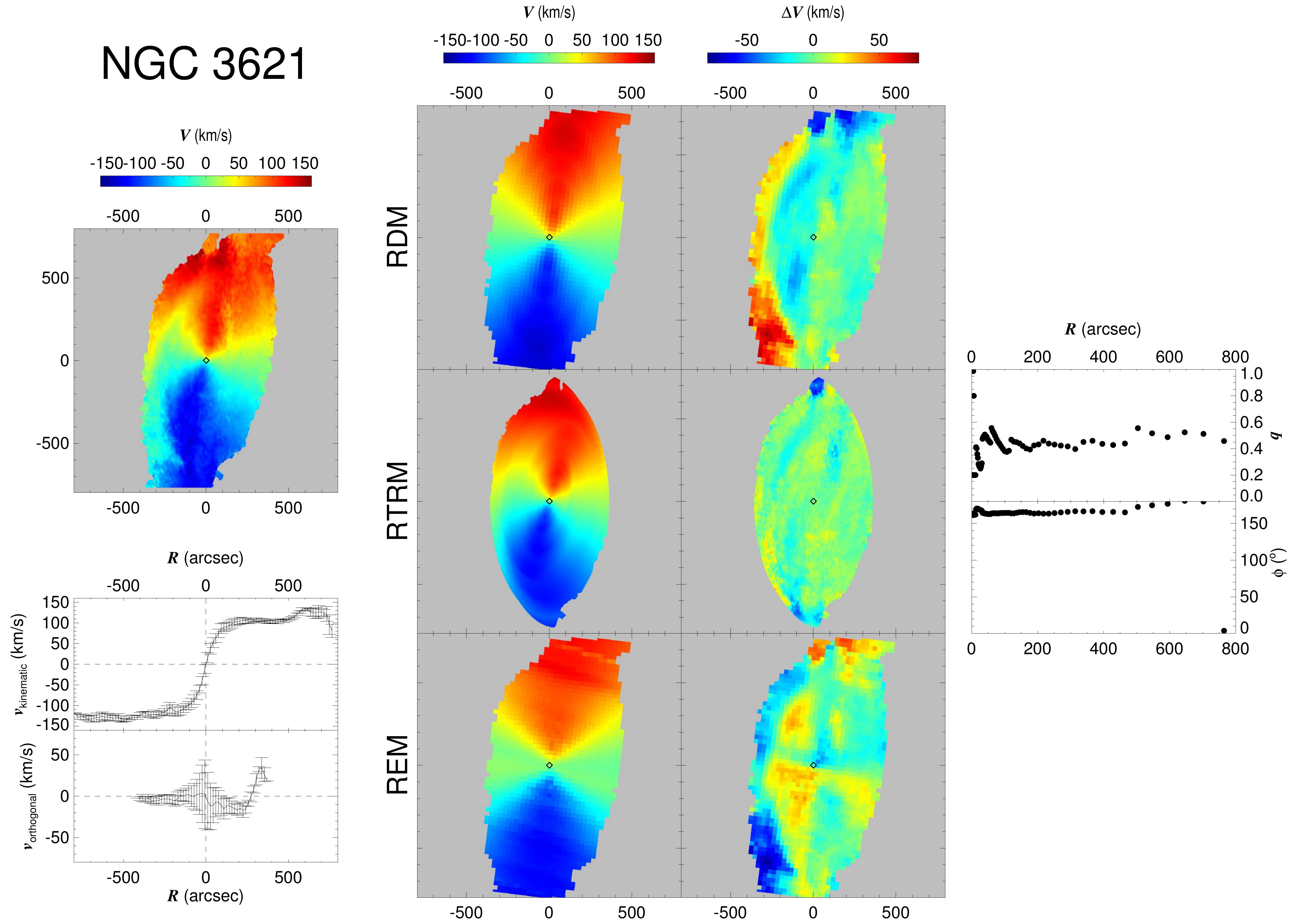}
\caption{As in Fig.\ref{ngc628} but for NGC 3621}
\label{ngc3621} 
\end{center}
\end{figure*}

\newpage
\clearpage

\begin{figure*}
\begin{center}
\includegraphics[width = 5in]{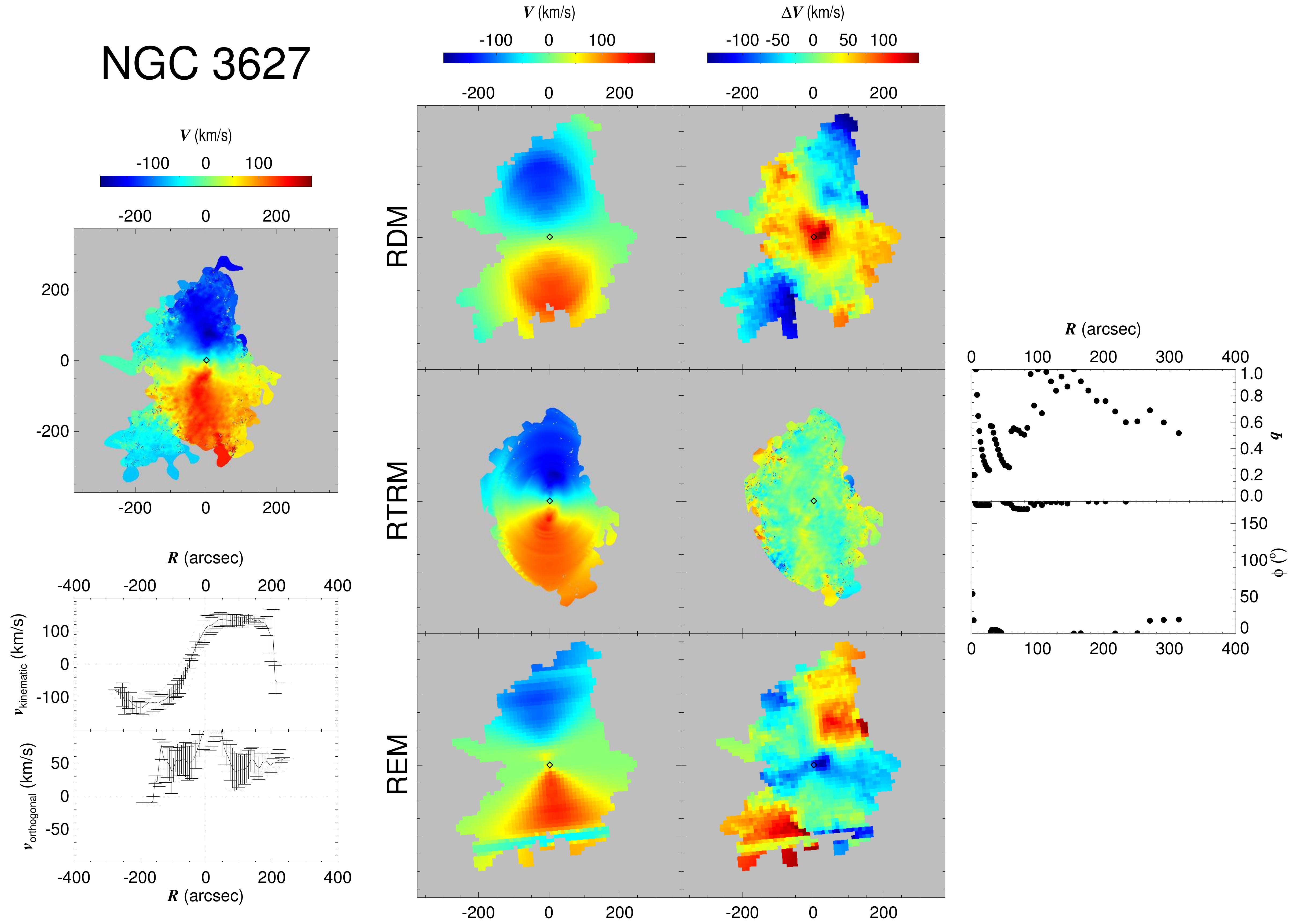}
\caption{SAs in Fig.\ref{ngc628} but for NGC 3627}
\label{ngc3627} 
\end{center}
\end{figure*}

\begin{figure*}
\begin{center}
\includegraphics[width = 5in]{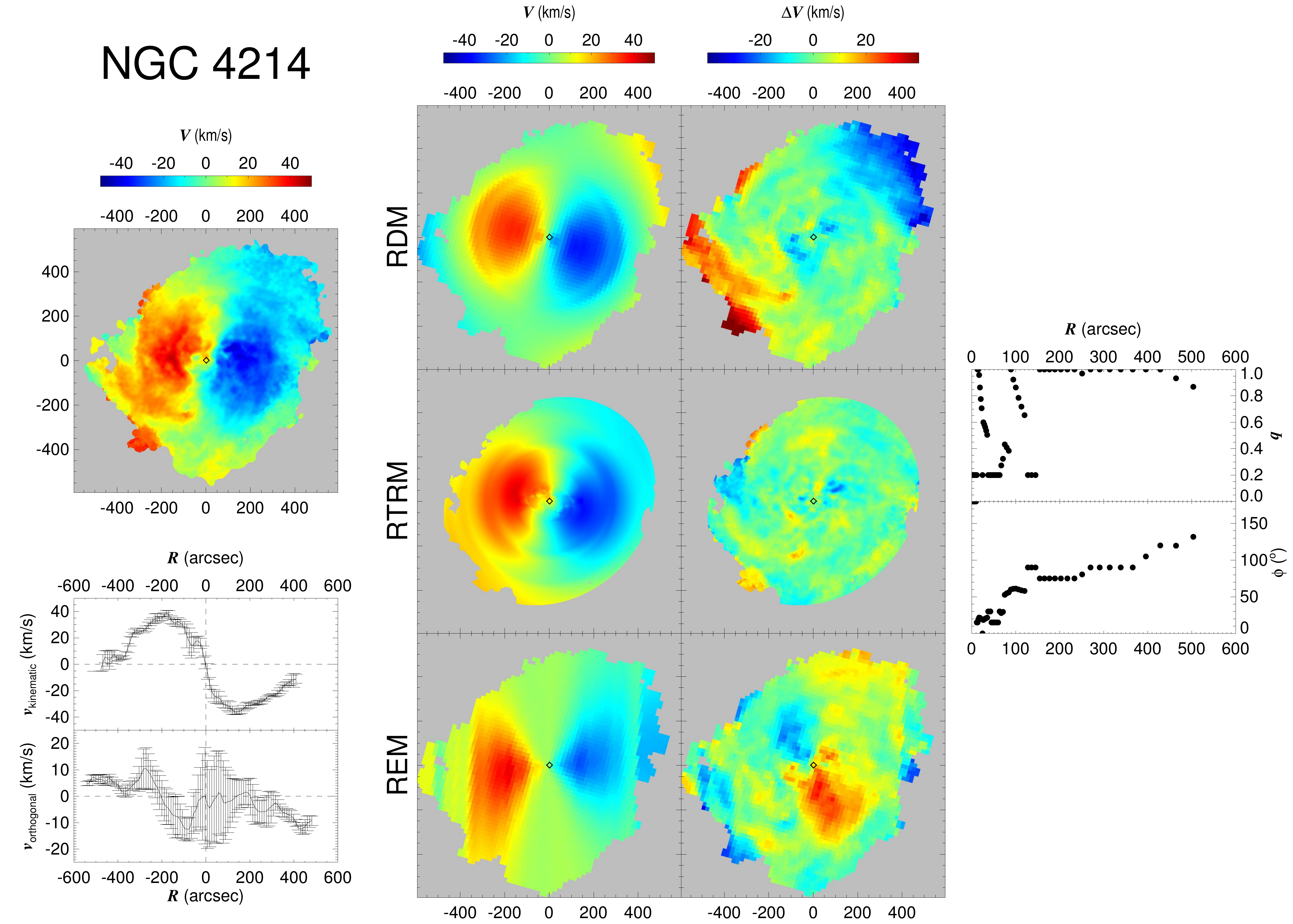}
\caption{As in Fig.\ref{ngc628} but for NGC 4214}
\label{ngc4214} 
\end{center}
\end{figure*}

\newpage
\clearpage

\begin{figure*}
\begin{center}
\includegraphics[width = 5in]{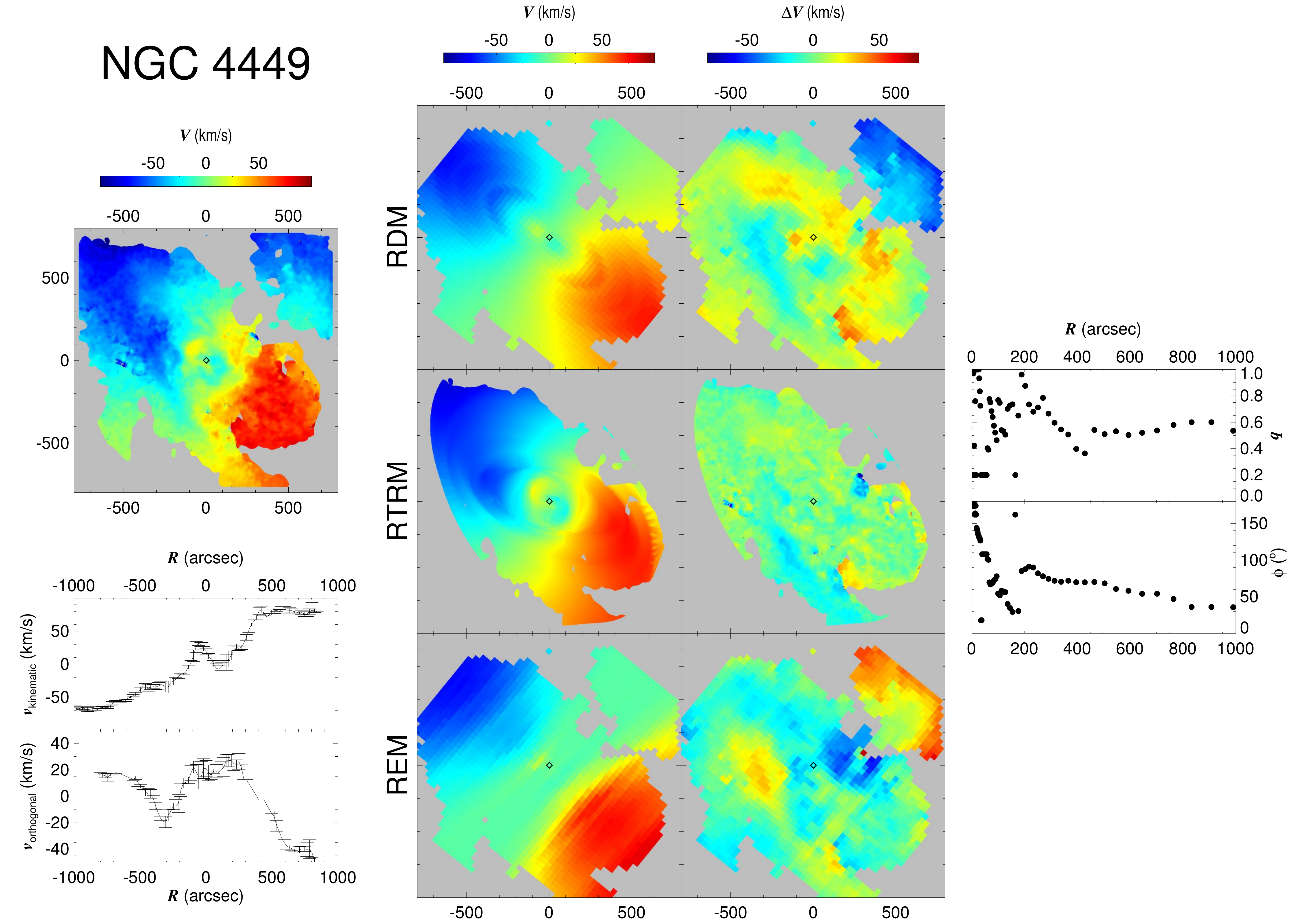}
\caption{As in Fig.\ref{ngc628} but for NGC 4449}
\label{ngc4449} 
\end{center}
\end{figure*}

\begin{figure*}
\begin{center}
\includegraphics[width = 5in]{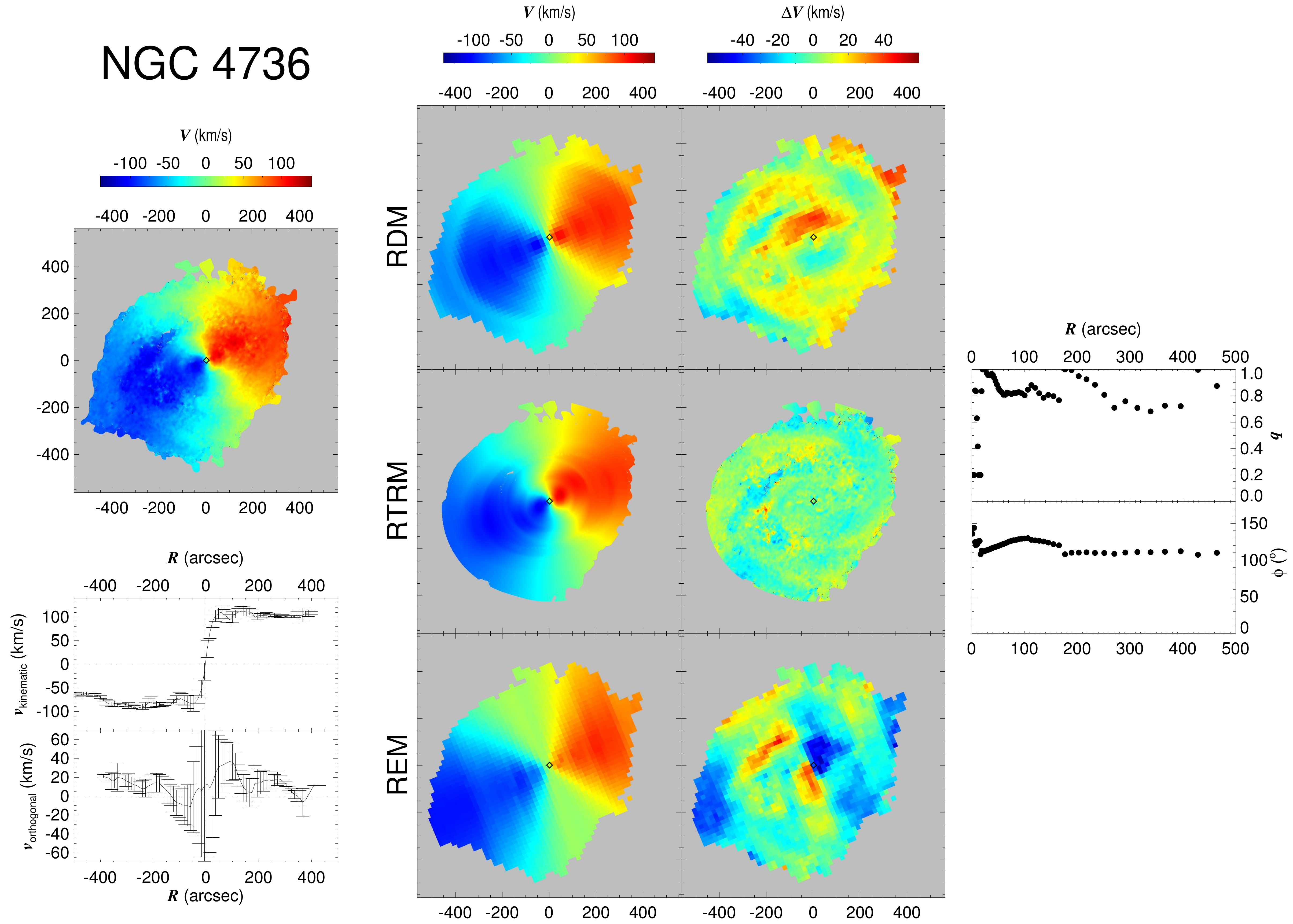}
\caption{As in Fig.\ref{ngc628} but for NGC 4736}
\label{ngc4736} 
\end{center}
\end{figure*}

\newpage
\clearpage

\begin{figure*}
\begin{center}
\includegraphics[width = 5in]{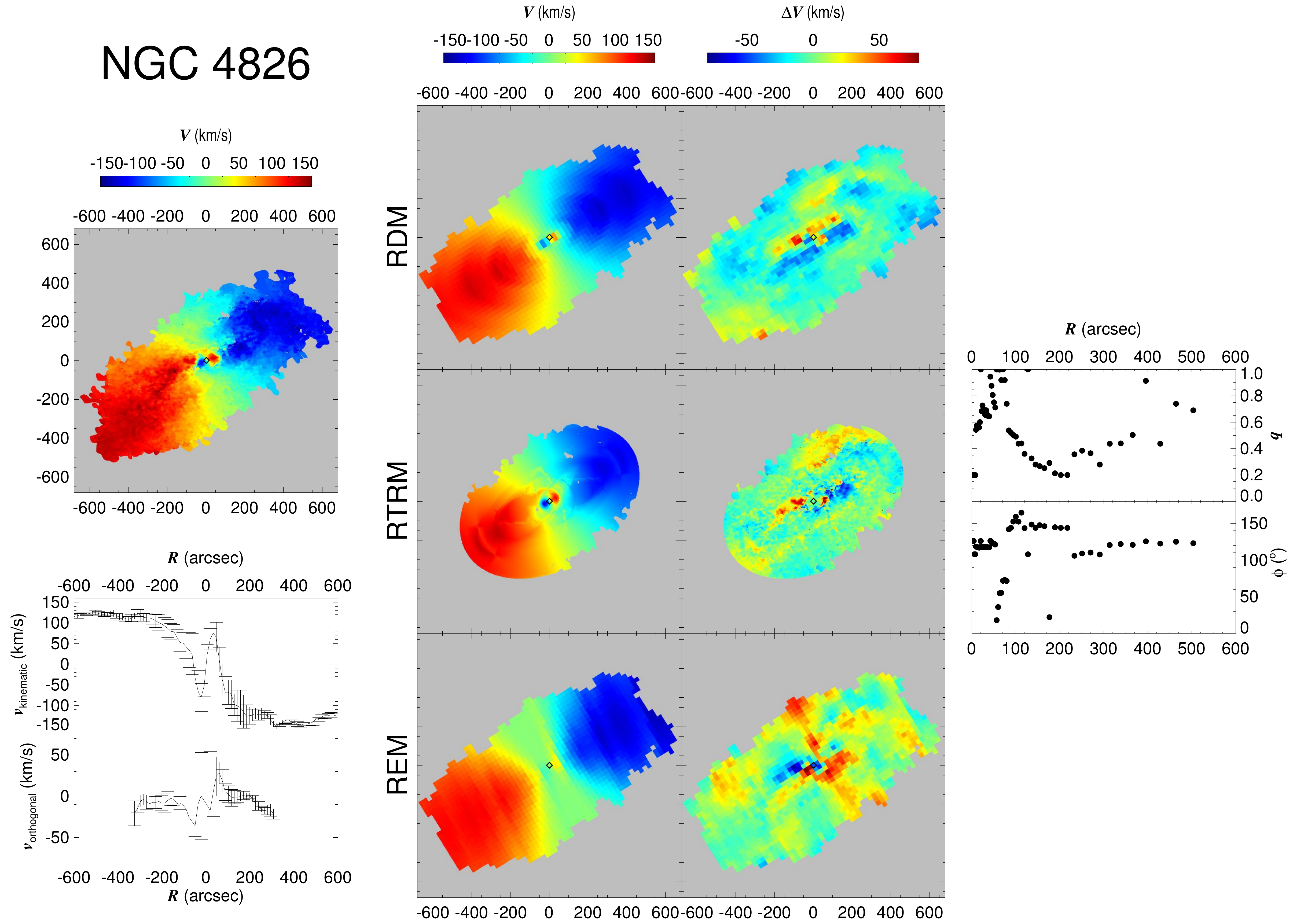}
\caption{As in Fig.\ref{ngc628} but for NGC 4826}
\label{ngc4826} 
\end{center}
\end{figure*}

\begin{figure*}
\begin{center}
\includegraphics[width = 5in]{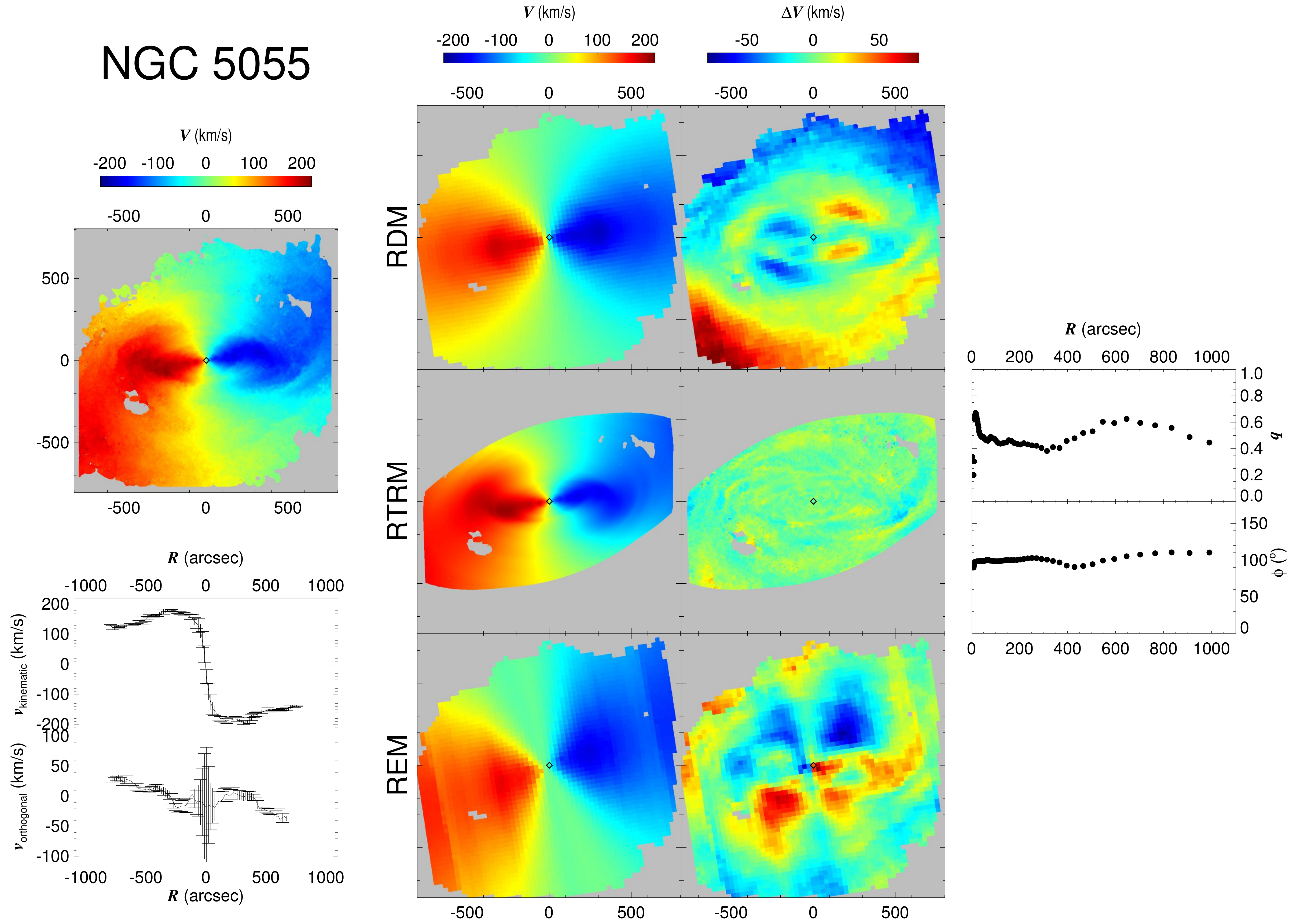}
\caption{As in Fig.\ref{ngc628} but for NGC 5055}
\label{ngc5055} 
\end{center}
\end{figure*}

\newpage
\clearpage

\begin{figure*}
\begin{center}
\includegraphics[width = 5in]{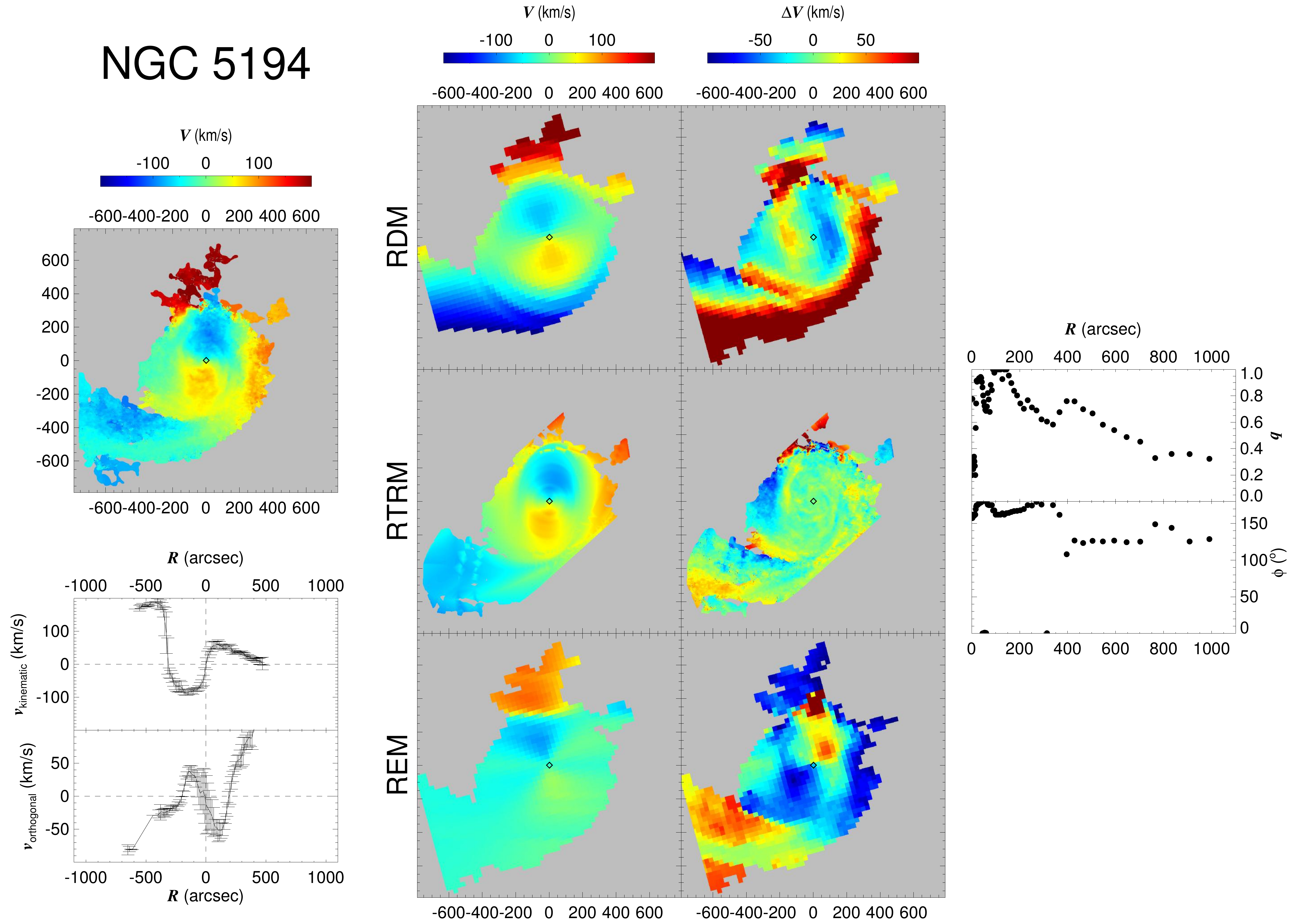}
\caption{As in Fig.\ref{ngc628} but for NGC 5194}
\label{ngc5194} 
\end{center}
\end{figure*}

\begin{figure*}
\begin{center}
\includegraphics[width = 5in]{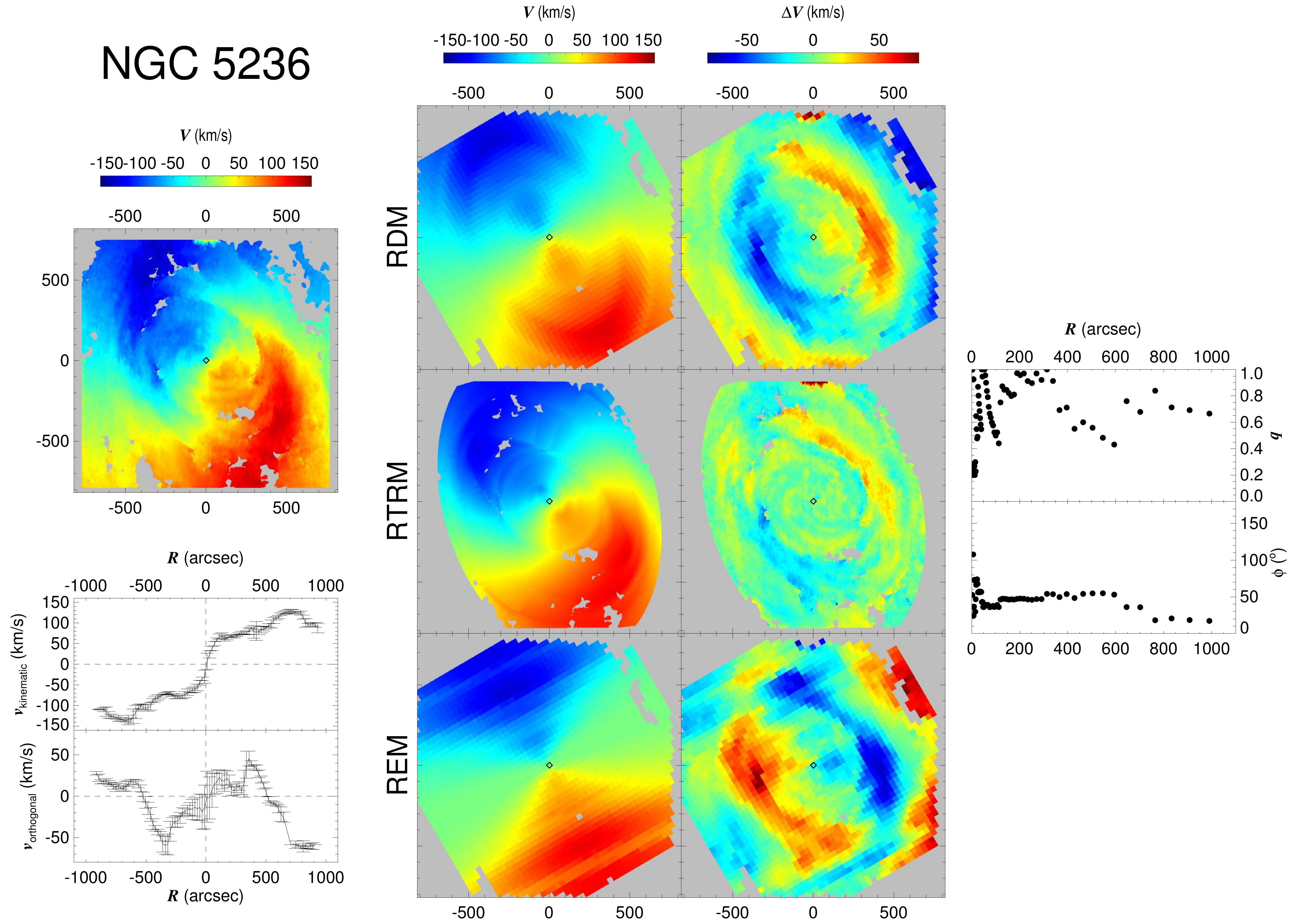}
\caption{As in Fig.\ref{ngc628} but for NGC 5236}
\label{ngc5236} 
\end{center}
\end{figure*}

\newpage
\clearpage

\begin{figure*}
\begin{center}
\includegraphics[width = 5in]{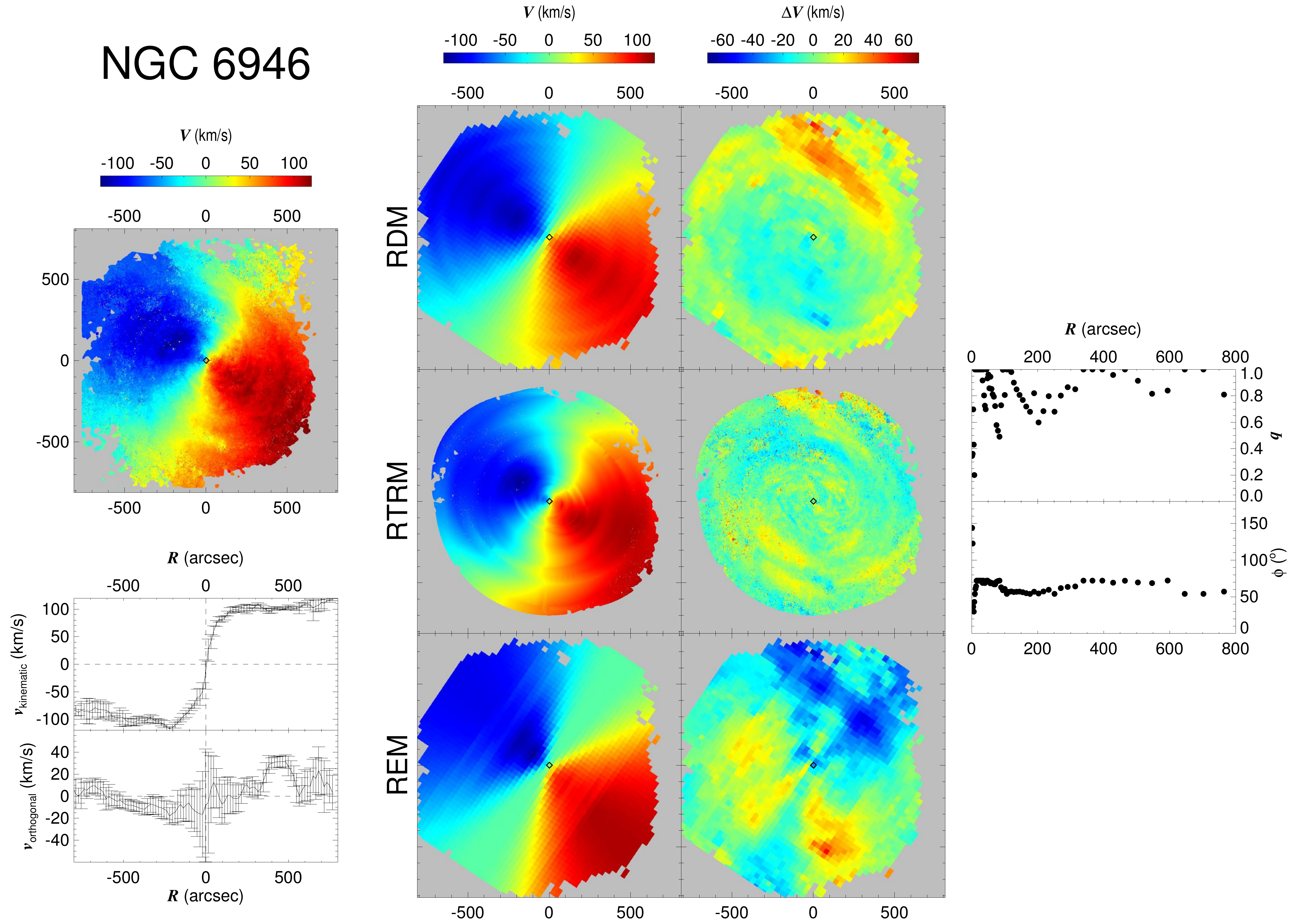}
\caption{SAs in Fig.\ref{ngc628} but for NGC 6946}
\label{ngc6946} 
\end{center}
\end{figure*}

\begin{figure*}
\begin{center}
\includegraphics[width = 5in]{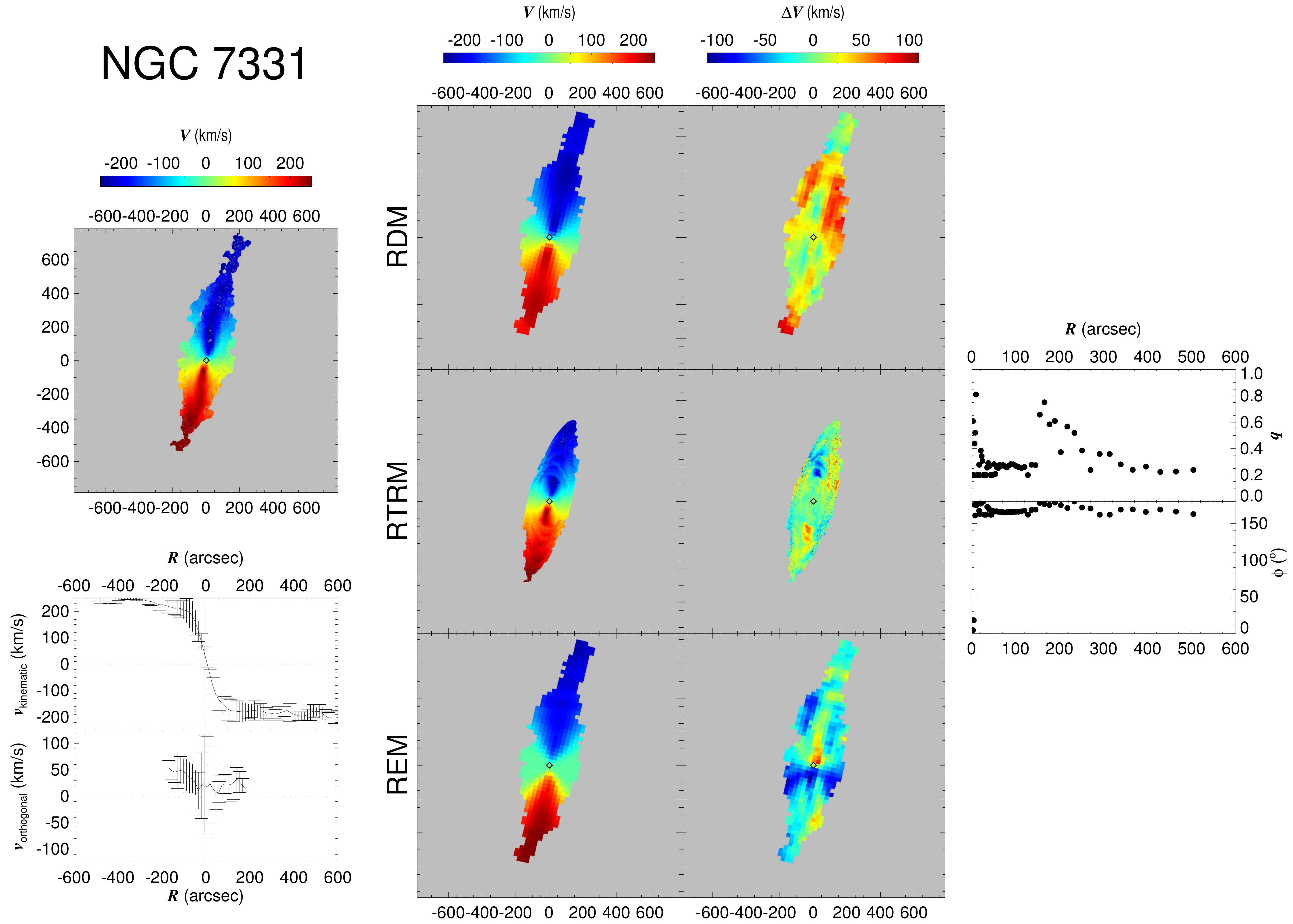}
\caption{As in Fig.\ref{ngc628} but for NGC 7331}
\label{ngc7331} 
\end{center}
\end{figure*}

\newpage
\clearpage

\begin{figure*}
\begin{center}
\includegraphics[width = 5in]{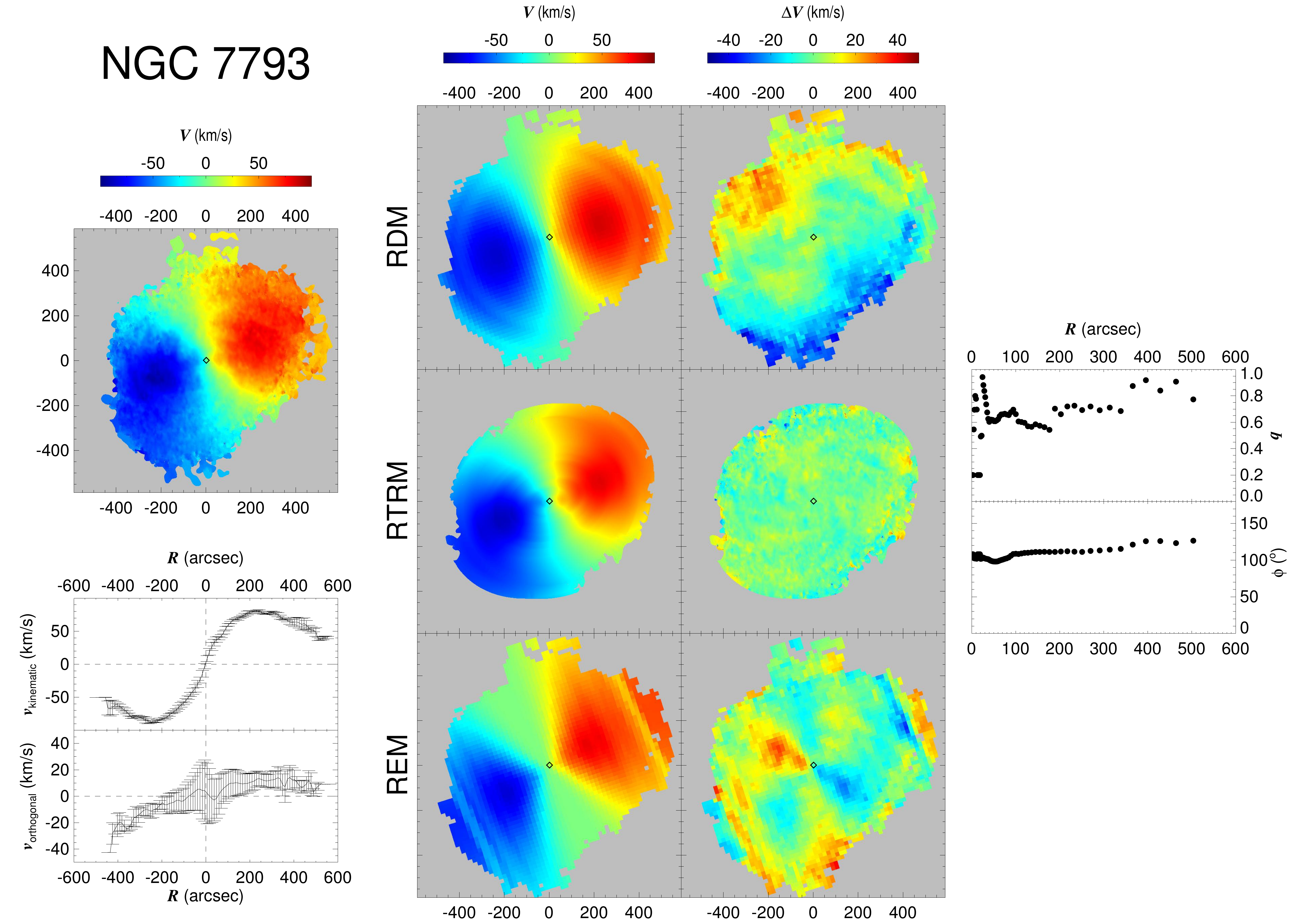}
\caption{As in Fig.\ref{ngc628} but for NGC 7793}
\label{ngc7793} 
\end{center}
\end{figure*}

\end{document}